\documentclass[a4paper]{aa}
\usepackage{epsfig,times}
\usepackage{color}
\usepackage{graphicx}
\usepackage{natbib}

\begin{document}

\title{Trapping Solids at the Inner Edge of the Dead Zone: 3-D Global
  MHD Simulations}
\author{Natalia Dzyurkevich$^1$, Mario Flock$^1$, Neal J. Turner$^{1,2}$,
          Hubert Klahr$^1$ and Thomas Henning$^1$}

\institute{
  $^1$Max Planck Institute for Astronomy, K\"onigstuhl 17, 
  69117 Heidelberg, Germany\\
  $^2$Permanent address: Jet Propulsion Laboratory, California
  Institute of Technology, Pasadena, California 91109, USA}
\offprints{natalia@mpia.de}

\date{\today}
\authorrunning{Dzyurkevich et al.}
\titlerunning{Trapping Solids in 3-D Global MHD Simulations}

\abstract{The poorly-ionized interior of the protoplanetary disk or
  `dead zone' is the location where dust coagulation processes may be
  most efficient.  However even here, planetesimal formation may be
  limited by the loss of solid material through radial drift, and by
  collisional fragmentation of the particles.  Both depend on the
  turbulent properties of the gas.}
{Our aim here is to investigate the possibility that solid particles
  are trapped at local pressure maxima in the dynamically evolving
  disk.  We perform the first 3-D global non-ideal
  magnetohydrodynamical (MHD) calculations of a section of the disk
  treating the turbulence driven by the magneto-rotational instability
  (MRI).}
{We use the ZeusMP code with a fixed Ohmic resistivity distribution.
  The domain contains an inner MRI-active region near the young star
  and an outer midplane dead zone, with the transition between the two
  modeled by a sharp increase in the magnetic diffusivity.}
{The azimuthal magnetic fields generated in the active zone oscillate
  over time, changing sign about every 150~years.  We thus observe the
  radial structure of the `butterfly pattern' seen previously in local
  shearing-box simulations.  The mean magnetic field diffuses from the
  active zone into the dead zone, where the Reynolds stress
  nevertheless dominates, giving a residual $\alpha$ between $10^{-4}$
  and $10^{-3}$.  The greater total accretion stress in the active
  zone leads to a net reduction in the surface density, so that after
  800~years an approximate steady state is reached in which a local
  radial maximum in the midplane pressure lies near the transition
  radius.  We also observe the formation of density ridges within the
  active zone.}
{The dead zone in our models possesses a mean magnetic field,
  significant Reynolds stresses and a steady local pressure maximum at
  the inner edge, where the outward migration of planetary embryos and
  the efficient trapping of solid material are possible.}

\maketitle

%---------------------------------------------------------------------

\renewcommand{\vec}[1]{\mbox{\boldmath$#1$}}

\newcommand{\etal}{{\rm et al. }}

%---------------------------------------------------------------------

\section{Introduction}

Forming planets in a protoplanetary disk with a power-law surface
density profile is difficult for several reasons.  First, solid
material on accumulating into meter-sized boulders quickly spirals to
the star, transferring its orbital angular momentum to the gas
\citep{wei77,nak86,tak02,you04,bra07}.  Second, collisions between the
constituents lead to disruption rather than growth when rather low
speed thresholds are reached \citep{blu98,pop99,blu00}.  Bodies in the
meter size range are destroyed in impacts as slow as some cm/s
\citep{ben00}.  Turbulence in the gas readily yields collisions fast
enough to terminate growth \citep{bra08a}.  Third, Earth-mass
protoplanetary cores are prone to radial migration resulting from the
tidal interaction with the gas, and in the classical type~I migration
picture quickly migrate all the way to the host star
\citep{gol80,war86}.

All three of these problems could be solved by the presence of local
radial gas pressure maxima, which trap the drifting particles
\citep{hag03}, leading to locally enhanced number
densities and high rates of low-speed collision
\citep{lyr08,bra08b,kre09}.  With sufficient local enhancement, one
can envision the direct formation of planetesimals via collapse under
the self-gravity of the particle cloud, bypassing the size regime most
susceptible to the radial drift.  Furthermore, the radial migration of
Earth-mass protoplanets can be slowed or stopped by varying the
surface density and temperature gradients \citep{mas06}.  Migration
substantially slower than in the classical picture appears to be
required to explain the observed exoplanet population under the
sequential planet formation scenario \citep{sch09}.

The formation of local pressure maxima is governed by the radial
transport of gas within the disk.  The magneto-rotational instability
or MRI \citep{bal91,bal98} is currently the best studied candidate to
drive such flows.  Local shearing-box calculations show that the
instability leads to long-lasting turbulence and to angular momentum
transfer by magnetic forces, provided the magnetic fields are
well-coupled to the gas \citep{haw95,bra95,san04,joh09}. Global ideal 
 MHD calculations have been performed in various astrophysical
contexts: protoplanetary disks \citep{ste01,arl01,fro05,fro06,fro09},
black hole accretion toruses \citep{haw00}, and galactic disks
\citep{dzy04}.  All the global simulations included neither a physical
magnetic diffusivity nor a physical viscosity.  Meanwhile, from local
shearing-box studies it is known that the strength of the saturated
MRI turbulence depends critically on the resistivity and viscosity
\citep{les07,fro07a,fro07b}.  In particular, whereas the molecular
viscosity is small in protostellar disks, the gas is so weakly ionized
in the cold disk interior \citep{gam96,ige99,sem04} that the Ohmic
resistivity shuts down the linear MRI and prevents the development of
turbulence \citep{san98,san02a,san02b,fle03,tur07}.

In this paper we present the first global resistive MHD calculations
to include the `dead zone' where the rapid diffusion of the magnetic
fields prevents magnetorotational turbulence.  Local pressure maxima
form in the calculations in two ways: at dead zone edges and in zonal
flows.  The dead zone edges yield long-lived rings of enhanced surface
density near locations where a gradient in the ionization fraction
leads to a jump in the accretion stress \citep{kre07}.  The zonal
flows on the other hand result from local fluctuations in the Maxwell
stress in the turbulence, and lead to pressure maxima with lifetimes
of a few orbits \citep{joh09}.

In the next section we describe our disk model and the choice of
magnetic resistivity profiles.  The third section presents the
results, starting from the global properties of the magnetic field,
followed by the effects of the resistivity jump on the surface
density.  In the fourth section we discuss the interaction of the
magnetic fields with the density rings and with radial minima
appearing in the turbulent activity.  Our main results are summarized
in section~5.

\begin{figure}[h]
\begin{center}
 \includegraphics[width=3.2in]{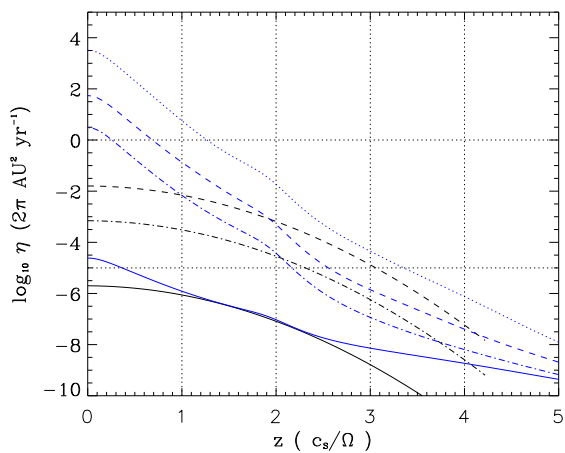}
 \caption{ Vertical profiles of magnetic diffusivity.  Black lines
   show the profiles adopted for simulations of the dead zone, with
   $\eta_{0}=2\cdot{}10^{-6}$ (solid), $\eta_{0}=7\cdot{}10^{-4}$
   (dot-dashed) and $\eta_{0}=0.016$ (dashed) (see Eq.~\ref{eta}).
   Blue lines show the estimations of magnetic diffusivity made for
   the disk at 4.5 AU using the simple gas-phase reaction set
   \citep{opp74,ilg06} together with dust grains.  The solid line is
   for no grains, dotted for $0.1 {\rm \mu m}$, dashed for $1 {\rm \mu
     m}$ and dot-dashed for $10 {\rm \mu m}$ dust grains.}
\label{fig1-0}
\end{center}
\end{figure}

\section{Model Description}

The initial setup for the models is very similar to those studied by
\citet{fro06} for the ideal MHD case.  In our models, MRI-driven
turbulence is operating in locally-isothermal disk, with a fixed
spatial distribution for the temperature.  To describe the dead zone
in the protoplanetary disk, we include the Ohmic dissipation in our
models.  The Ohmic dissipation is the largest non-ideal term in the
induction equation under typical dusty conditions \citep{war07}.
Ambipolar diffusion and Hall effect are not considered for the sake of
simplicity.  The vertical profile of magnetic diffusivity (see section
2.1) is adopted from separate chemistry calcullations and is fixed in
space and time.

We solve the set of MHD equations using 3D global simulations on a
spherical grid $(r, \Theta, \phi)$,
\begin{equation}
\frac{\partial \vec{B}}{\partial t}=
\nabla \times [\vec{u}\times \vec{B}-\eta(\Theta)\nabla \times\vec{B}],
\end{equation}
\begin{equation}
\frac{\partial \vec{u}}{\partial t}=
-\frac{1}{\rho}\nabla P +  \nabla \Psi + \frac{1}{4\pi\rho}[\nabla \times\vec{B}]\times\vec{B},
\end{equation}
\begin{equation}
\frac{\partial\rho}{\partial t} + \nabla \cdot(\rho\vec{u} )=0.
\end{equation}
%--%----------
The notation is the usual one. $\Psi$ is a point-mass gravitational
potential and $\sqrt{G{M}_{\star}}$ is set to unity in the code units.
We use the locally isothermal approach to describe vertically
stratified disks, adopting $P=c_{\rm s}^2(r) \rho(r, \Theta)$ with a
sound speed $c_{\rm s}=c_{0}/(r\cdot{}\sin(\Theta))$ and $H/R=0.07$.
The initial field setup is exactly the equilibrium solution,
\begin{equation}
u_{\phi}=\frac{1}{r} \sqrt{1-\frac{c_{0}^2(1+a)}{ \sin{\Theta}} },  
\label{6ia}
\end{equation}
\begin{equation}
\rho=\rho_{0} \frac{1}{(r\sin{\Theta})^a} \exp{\left( -\frac{\cos{\Theta}^2
    }{2c_{0}^2\sin{\Theta}^2 }  \right)},
\label{7ia}
\end{equation}
where  $a=3/2$.

We consider the inner part of the protoplanetary disk, which is
heavier and warmer compared to a minimum solar-nebula.  In order to
mimic a 'dead zone' and the ionization thresholds, we adopt fixed
magnetic diffusivity profiles $\eta(\Theta)$.  We use the estimations
from dynamical disk chemistry simulations for the adopted disk
parameters $\Sigma, T$ \citep{tur07}.  The surface density is
$\Sigma=(R/3.75{ \rm AU})^{-1/2} \cdot 1700\ \rm g/cm^2$, and the
temperature is $T=(R/3.75{ \rm AU})^{-1} \cdot 280 \rm K$, where $R$
is cylinder radius.  The units are $[u_{\phi}]=2\pi\ {\rm AU/yr}=29.8\
{\rm km/s}$ for velocity and $[\eta]=2\pi\ {\rm
  AU^2/yr}=4.47\cdot{}10^{19}\ {\rm cm^2/s}$ for magnetic diffusivity.
The models are listed in Table~\ref{tab1}.

Our models include the disk part from 2 to 10 AU.  A purely azimuthal
magnetic field is chosen as seed field for MRI turbulence, which is
$P_{\rm gas}/P_{\rm mag}=25$ everywhere in the disk. The Alfv\'en
limiter of $c_{\rm lim}=14 c_{0}$ is applied.  We use reflective
radial boundary condition, with buffer zones applied at radii $2{\rm
  AU}<r<2.5 {\rm AU}$ and $9.5{\rm AU}<r<10 {\rm AU}$.  We apply the
magnetic diffusivity within the buffer zones: $\eta_{\rm buffer}$ is
$10^{-5}$ at the radial boundary and decreases linearly towards the
physical domain.  The buffer zones have both damping of radial
velocity towards zero at the border, and diffusing away the magnetic
eddies approaching the radial boundary.

Periodic boundary conditions are applied both for azimuthal and
vertical (i.e. $\Theta$) domain borders.

%%%%%%%%%%%%%%%%%%%%%%%%%% table
\begin{table*}
\begin{center}
  \caption{Model properties and midplane $\alpha$-stresses inside
    ($\alpha_{\rm A}$) and outside ($\alpha_{\rm D}$) of the
    ionization threshold radius $r_{\rm th}$ }
%  The domain size is fixed
%   at [8 AU : $8.4 H$: $\pi/4$].  
%   The time duration of each model is
%    given in years, and the mark $*$ is given when the steady-state
%   has not been reached.  }
\label{tab1}
{\scriptsize
\begin{tabular}{|l|c|c|cc|cc|c|}\hline
{\bf Model}  & {\bf Resolution } & $r_{\rm th}$[AU] & {\bf
  $\eta_{\rm A}$} & {\bf  $\eta_{\rm D}$} 
& {\bf $\alpha_{\rm A}$} & {\bf  $\alpha_{\rm D}$} & Time [years]\\
\hline
$\rm M_{IR}$  & [256:128:64] & 4.5 & $2.10^{-6}$ & $7.10^{-4}$ &
  $1.9 \cdot{}10^{-3}$ & $1.0\cdot{}10^{-3}$ & 880 \ \\
$\rm M_{IR/2}$  & [128:64:32]& 4.5 & $2.10^{-6}$ & $7.10^{-4}$ &
  $ 1.3\cdot{}10^{-3}$ & $ 2.1 \cdot{}10^{-4}$ & 960 \ \\
$\rm M_{ID}$  & [256:128:64] & 4.5& $2.10^{-6}$ & $0.016$
& $ 1.6 \cdot{}10^{-3}$ & $ 2.9\cdot{}10^{-4}$ & 650 \ \\
$\rm M_{IR6}$  & [256:128:64]& 6.5 & $2.10^{-6}$ & $7.10^{-4}$ &
  $2.9\cdot{}10^{-3}$ & $1.6\cdot{}10^{-3}$ & 1100 \\
$\rm M_{\rm ideal}$ & [256:128:64]& none &  $2.10^{-6}$
&  $2. 10^{-6}$ & $5.4\cdot{}10^{-3} $ & $5.4\cdot{}10^{-3} $ & 682 $*$ \\
\hline
\end{tabular}
}
 \end{center}
\vspace{1mm}
 \scriptsize{
}
\end{table*}

\subsection{Ionization Thresholds and Influence of Dust Grains}

An estimate of the magnetic diffusivity vs. height at 4.5 AU is shown
in Fig.~\ref{fig1-0}.  The midplane diffusivity with dust grains
appears to be substantially higher then it is possible to include in
the MHD simulations.  The four blue curves from top to bottom are
demonstrating the magnetic diffusivity in code units for the gas and
dust grains of 0.1, 1 and 10 microns, and no grains.  We have used the
simple gas-phase reaction set of \citet{opp74} together with the grain
surface chemistry of \citet{ilg06} for the classical dust to gas
ratio.  Ionization by stellar X-rays, cosmic rays and long-lived
radionuclides is included.  The penetration depths are assumed $8\rm
g/cm^2$ for the X-rays and $96$ for the cosmic rays.

The exact calculations of chemistry and dust behavior in the thermally
evolving global disk is a hard task with many free parameters.  After
planetesimals form, the dust mass fraction will be lower than the
interstellar value.  We shall bear in mind that CRP stopping depth can
be as low as $36\rm g/cm^2$ \citep{gla09}.

Here we simplify the situation and adopt the following
time-independent vertical profile of magnetic diffusivity,
\begin{equation}
\eta=\eta_{0}
\exp{\left(\frac{\sin{\Theta}-1}{c_{0}^2}  \right)^{1.55}},
\label{eta}
\end{equation}
where $\eta_{0}$ is the midplane value of magnetic diffusivity.  In
Fig.~\ref{fig1-0}, black lines show the profiles adopted for our
simulations, with $\eta_{0}=2\cdot{}10^{-6}$ (solid),
$\eta_{0}=7\cdot{}10^{-4}$ (dot-dashed) and $\eta_{0}=0.016$ (dashed).
To simulate the situation mentioned in \citet{kre07,kre08}, we drop
the radial dependence of $\eta$ and suggest that $\eta_{0}$ makes the
jump at the chosen threshold radius $r_{\rm th}$. For example, in
model $\rm M_{IR}$ (section 2.2, Table 1), the Ohmic diffusion is
following a black solid line in Fig.~\ref{fig1-0} for radii from 2 to
4.5 AU, and a dot-dashed line for radii from 4.5 to 10 AU.  The
magnetic diffusivity in the disk without dust grains is low, so that
we have MRI-active region from 2 to 4.5 AU.  The dead zone begins
where the dust grains are present.  The exact location of the inner
edge of the dead zone (i.e. $r_{\rm th}$) depends on the properties of
the star and the dust in the disk, and can vary from object to object.
Our choice for locating the threshold $r_{\rm th}$ is quite arbitrary.
We place $r_{\rm th}$ roughly in the middle of our inner global disk
patch (Table~\ref{tab1}).

\begin{figure*}[ht]
\begin{center}
\includegraphics[width=5.5in]{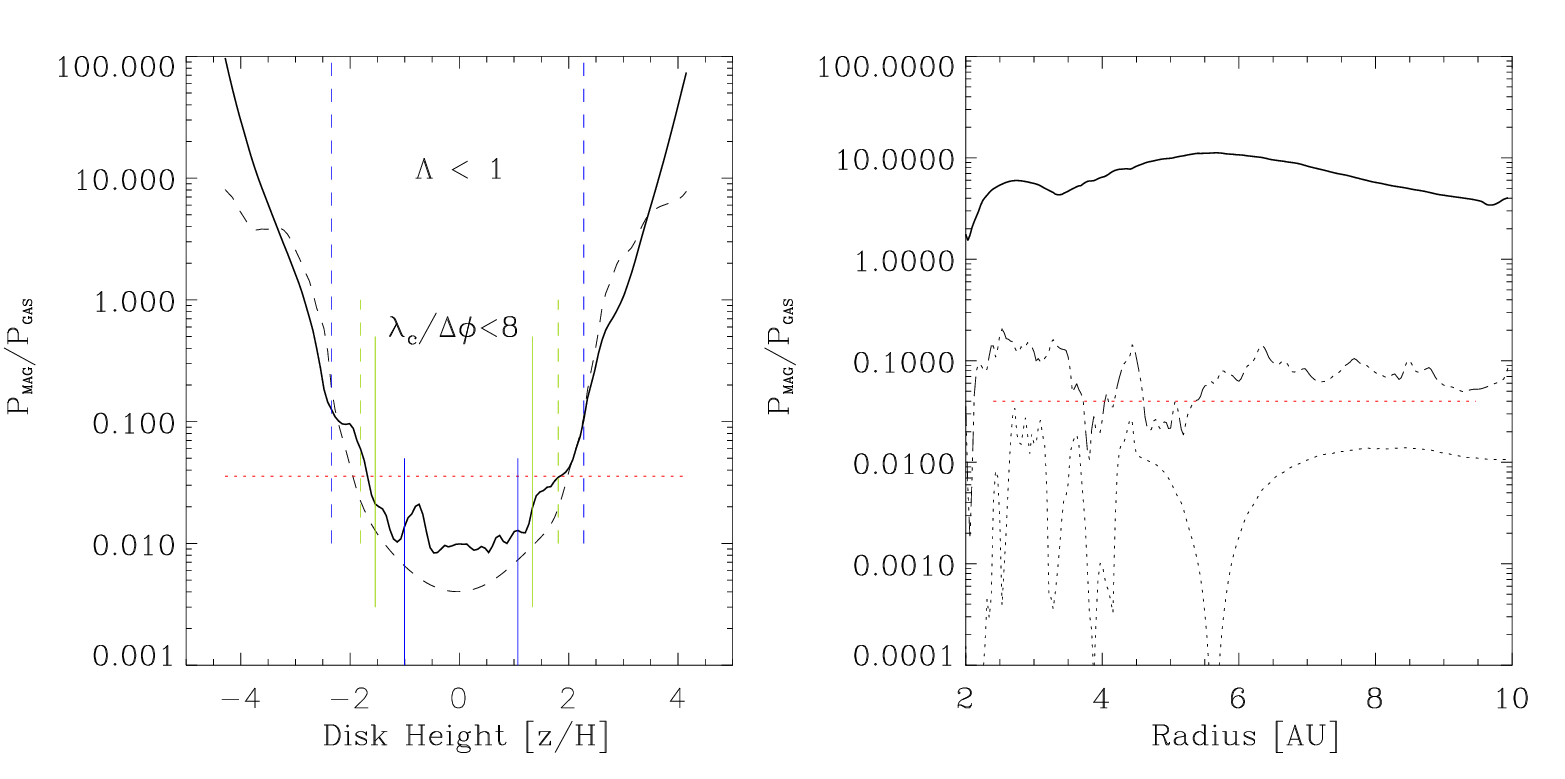}
\includegraphics[width=5.5in]{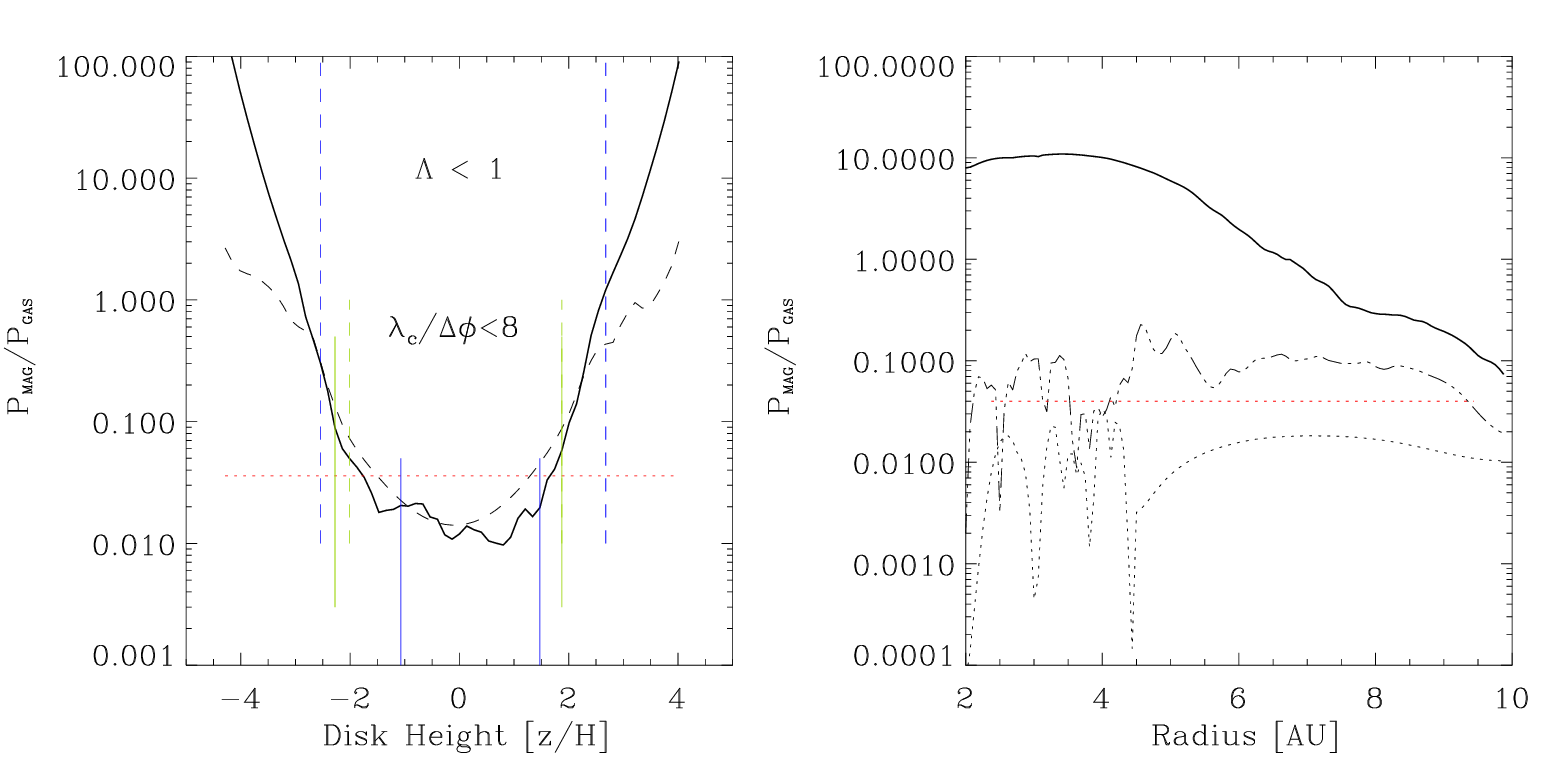}
\caption{ Inverse plasma $\beta$. Top: model $\rm M_{IR}$ at $t=300$
  years; Below: $\rm M_{IR/2}$ (halved space resolution) at the same
  time.  Left: The change of inverse plasma $\beta$ vertical
  distribution from initial (red) to the convex shape (black lines).
  The solid line stands for inverse plasma beta averaged within active
  zone ($2.5\to 4.5$ AU).  Dashed line is for the $1/\beta$ averaged
  over the patch of the dead zone ($5\to 7$AU).  Vertical bars mark
  the disk height where the Els\"asser number drops below unity (blue
  lines, solid for active and dashed for dead zones). Green lines
  border the midplane region with $\lambda_{\rm c}/\Delta \phi<8$
  (Eq.~\ref{resol}).  Right: Radial dependence of vertically averaged
  $1/\beta$ (solid line), $1/\beta$ at the midplane (dotted line) and
  at $z=2H$ (dashed line). }
\label{Bphi2}
\end{center}
\end{figure*}

\subsection{Set of Simulations}

Models in Table~\ref{tab1} have an ionization threshold posed either
at $4.5 \rm AU$ or $6.5 \rm AU$.  Midplane values for magnetic
diffusivity are noted in Table~\ref{tab1} as $\eta_{\rm A}$ and
$\eta_{\rm D}$ (`Active' and `Dead') for gas states inside and outside
the threshold radius.  
  The time duration of each model is
   given in years, and the mark $*$ is given when the steady-state
  has not been reached.
Vertical profiles for magnetic diffusivity
follow Eq.~\ref{eta} and are demonstrated in Fig.~\ref{fig1-0} with
black lines.  Each model combines two diffusivity profiles, except the
run $\rm M_{IDEAL}$.

In Table 1, notations are 'I' for quasi-ideal MHD state with
$\eta_{0}=2\cdot{}10^{-6}$ (Fig.~\ref{fig1-0}, solid black line), 'R'
for the gas disk with $10 {\rm\mu m}$-sized dust grains
($\eta_{0}=7\cdot{}10^{-4}$, dot-dashed black line), 'D' for the case
of $1\rm \mu m$ grains ($\eta_{0}=0.016$, dashed black line).  Our
adopted magnetic diffusivity profiles for the disk with $1\rm \mu m$
and $10 {\rm\mu m}$-sized dust grains will allow the turbulent MRI
layers beyond $3$H and $2$H, correspondingly.  The peak values of blue
curves in Fig.~\ref{fig1-0} are leading to unacceptable short time
steps in resistive MHD simulations.  We have observed that it is not
convenient to compute the regions with $\eta > 0.1 \rm AU^2/yr$ with
standard MHD codes, because of the dramatic shortening of the time
step.  For this numerical reason, we take the magnetic diffusivity
slightly different as the chemistry models predict.  Our adopted
profiles of magnetic diffusivity (black curves, Fig.~\ref{fig1-0})
allow to match the values of chemical models at 2 AU and remain above
the numerical dissipation for region between 2H and 3H.  Reducing of
the magnetic diffusivity in the dead zone may influence how fast the
global magnetic fields are diffused into the dead zone, whereas the
MRI modes are damped all the same.

%%%%%%%%%%%%%%%%%%%%%%%%%%%%%%%%%%%%%%%

\subsection{Calculation of Turbulent Stresses}

An important outcome of our simulations is the magnitude of the
Reynolds and Maxwell stresses.  To calculate the latter, we use the
approach described in \citet{fro06} for curvi-linear coordinates.  The
turbulent viscosity can be described as 
$\nu=\alpha c_{\rm s}^2/\Omega$, 
where the main component of the $\alpha$ stress tensor
is
\begin{equation}
\alpha_{r,\phi}=\frac{T_{\rm M}+T_{\rm R}}{\langle{ P \rangle} },
\label{8}
\end{equation}
or $\alpha=\alpha_{\rm R}+\alpha_{\rm M}$.
The Reynolds and Maxwell stresses are calculated as
\begin{equation}
T_{\rm R} =  \langle{(\rho u_{\phi}^{'} u_{r}^{'}) \rangle},
\label{9}
\end{equation}
\begin{equation}
T_{\rm M} =  -\langle{(B_{\phi}^{'} B_{r}^{'})/4\pi \rangle},
\label{10}
\end{equation}
The mean pressure for azimuthal domain $\Delta \phi$ is
\begin{equation}
{\langle{ P(r) \rangle} }=c_{\rm s}(r)^2\Sigma(r) = \frac{c_{\rm s}(r)^2}{\Delta \phi} \int_{\Delta \Theta}\int_{\Delta \phi} \rho r 
 \sin{(\Theta)}  d\Theta d \phi.
\label{11}
\end{equation}

%%%%%%%%%%%%%%%%%%%%%%%%%%%%%%%%%%%%%%%%%%%%%%%%%%%%%%%%%%%%%%%%%%%%%%%%%%%

\section{Results}

In this section we describe our results and focus on two main issues.
First, we study the time evolution and radial dependence of magnetic
fields in our models.  Secondly we study the formation of long-lived
density rings which may or may not be able to trap solids in the disk
and thus trigger the onset of planet formation.  Table~\ref{tab1}
represents the set of models.  In 3.1 we discuss the issue of
resolution.  In 3.2 we describe the properties of global models with
the emphasis on the evolution of the magnetic fields.  In 3.3, the
radial behavior of resulting Maxwell and Reynolds stresses is
described.  In 3.4 we explore the evolution of the pressure rings in
time.  In 3.5 we demonstrate the change of rotation and the turbulent
properties of the gas in the rings and in the pressure bump at the
inner edge of the dead zone.
%%%%%%%%%%%%%%%%%%%%%%%%%%%%%%%%%%%%%%%%%%%%%%%%%%%%%%%%%%%%%%%%%%%%%%%%%%%%%%%%

\subsection{Azimuthal MRI and the Issue of Resolution}

The MRI from a purely azimuthal magnetic field (AMRI) leads to
non-axisymmetric perturbations.  The radial displacements of the
initial azimuthal field are enhanced due to the differential rotation.
This leads to the appearance of field components $B_{r}$ and turbulent
$B_{\phi}$.  The excess of magnetic pressure and the buoyancy lead to
the generation of the vertical magnetic field component.  The linear
analysis has been done in Balbus \& Hawley (1992).  The critical
wavelength for AMRI in units of the azimuthal grid size is
\begin{equation}
\lambda_{\rm c}/\Delta{\phi}=2\pi\sqrt{\frac{16}{15}{\frac{2}{\beta}}} c_{0} 
/  \Delta \phi ,
\label{resol}
\end{equation}
which follows from Eq. (15) in \citet{haw95}.  When
$\vec{k}\cdot\vec{V_{\rm A}} \sim \Omega$ and $|\vec{k}/k_{z}|$ is in
the right range, then the growth rate of the non-axisymmetric modes is
greatest.  In contrast to the magneto-rotational instability of
vertical magnetic field, the vertical wavenumber by AMRI is not
constant and increase with time during the linear stage of MRI.  The
largest total field amplification is expected for the modes with
largest possible $|k_{z}|$.  As a consequence, in numerical
simulations of AMRI it may be impossible to resolve all growing
wave-numbers and the total amplification is limited by the grid size
\citep{haw95}.  Nevertheless, the numerical study in \citet{haw95}
shows that for effective resolution above 8 grids per critical
wavelength in azimuthal direction the saturation of magnetic energy is
only weakly affected by the resolution.  Note, that the resolving of
azimuthal critical wavelength is important.  In Fromang \& Nelson
(2006), the resolution of more then 5 grids per wavelength has been
suggested as sufficient.

All our runs are made with the initial uniform plasma beta of 25.
Following Eq.~\ref{resol}, we have $\lambda_{\rm
  c}/\Delta{\phi}=10.47$ everywhere in the MRI-active disk for the
models with resolution of $[256:128:64]$.  Model $\rm M_{IR/2}$ with
halved resolution has $\lambda_{\rm c}/\Delta{\phi}=5.24$.  Note, that
the Ohmic dissipation poses an additional limitation for the excited
MRI wavelength.
%%%%%% Elsasser numbers

\begin{figure}[h]
\begin{center}
\includegraphics[width=3.1in]{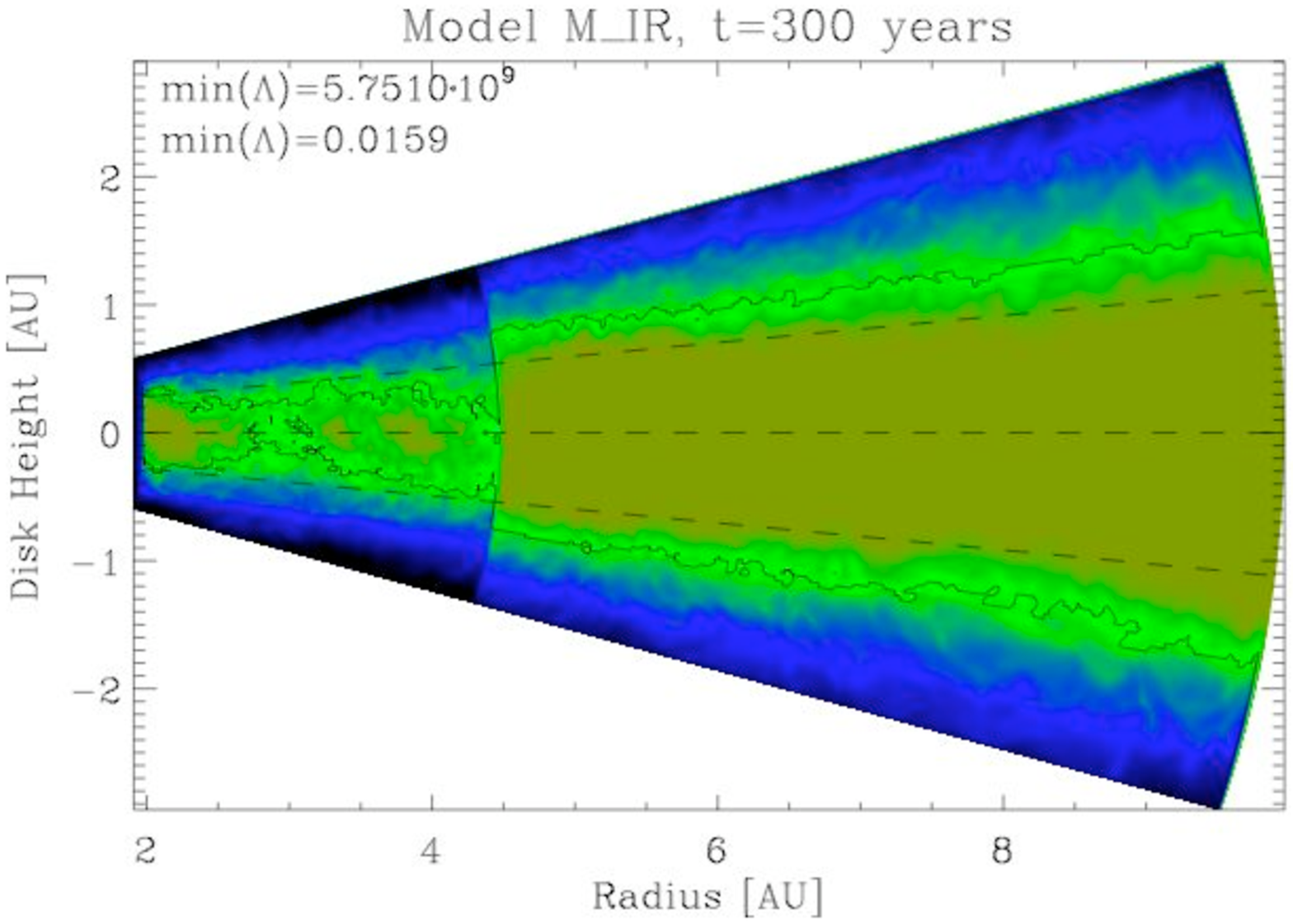}
\includegraphics[width=3.1in]{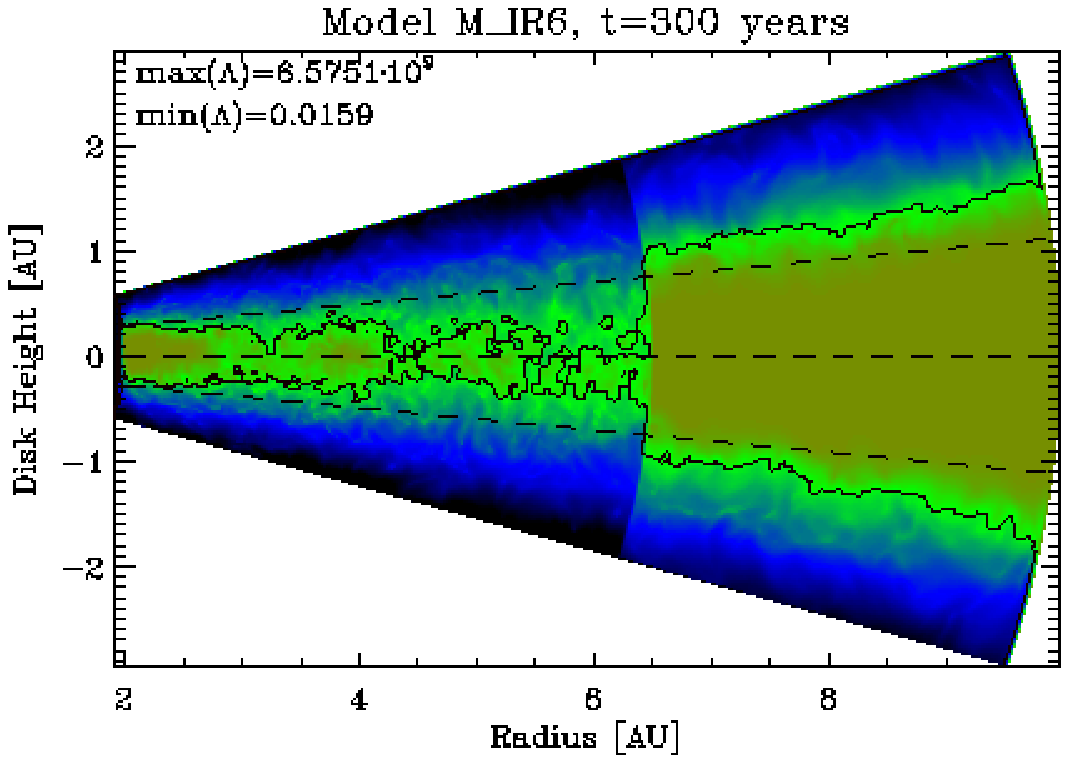}
\includegraphics[width=3.1in]{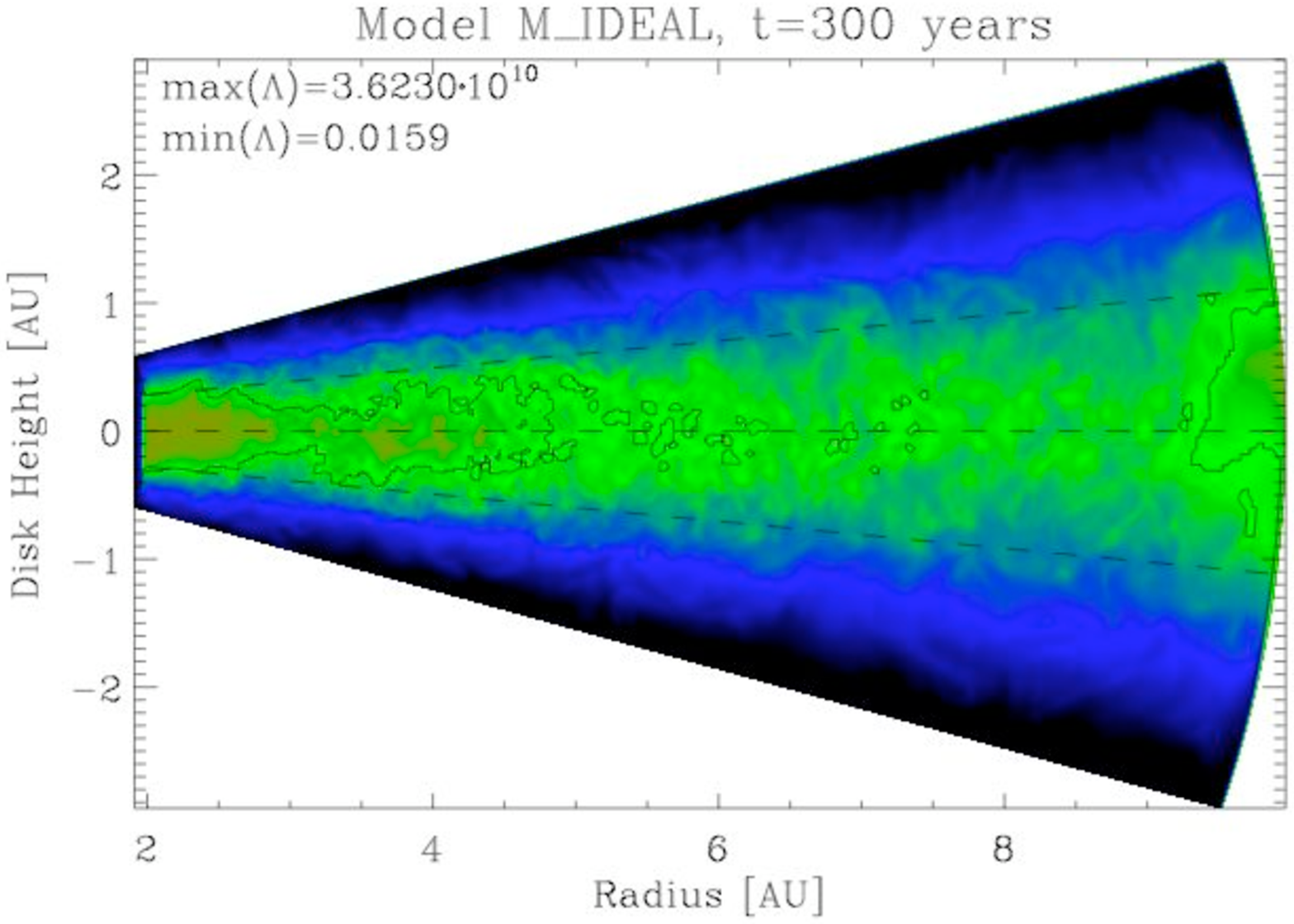}
\caption{ Els\"asser number $\Lambda=v_{{\rm A}z}^2/\eta\Omega$
  for models $\rm M_{IR}$ (top), $\rm M_{IR6}$ (middle) and $\rm
  M_{IR}$ (bottom).  The Els\"asser number is plotted in logarithmic
  color scale for edge-on view of the disk. Yellow-green color marks
  weak $B_{\Theta}$ magnetic fields which are stable to MRI.  Blue is
  marking the regions with the vertical magnetic field sufficiently
  strong to launch the MRI. The transition to stability at $v_{{\rm
      A}z}^2/\eta\Omega=1$ is indicated with a black line. Slices of
  turbulent magnetic fields are taken for $t=300$ years (orbits at 1
  AU) for each model. }
\label{mags}
\end{center}
\end{figure}

The effective resolution of $10.47$ is holding during the linear stage
of AMRI.  The azimuthal magnetic field is breaking into filaments of
opposite signs with $\lambda_{\Theta}<\lambda_{r}<<\lambda_{\phi}$ in
MRI-active zones.  The inverse plasma beta grows to 100 at the
midplane and the decreases below unity in the upper disk layers.  The
effect of expelling the magnetic field into the corona becomes visible
after 30 local orbits (section~3.2, Fig.~\ref{butter1}).  Note, that
in low-resolution tests with $\lambda_{\rm c}/\Delta{\phi}=2.62$ ($\rm
M_{IDEAL}$ with resolution of $r:\Theta:\phi=[64:16:8]$, excluded from
Table 1), there is no MRI exited and the azimuthal magnetic field
remains in its initial shape for few hundred years.

\begin{figure}[ht]
\begin{center}
\includegraphics[width=3.0in]{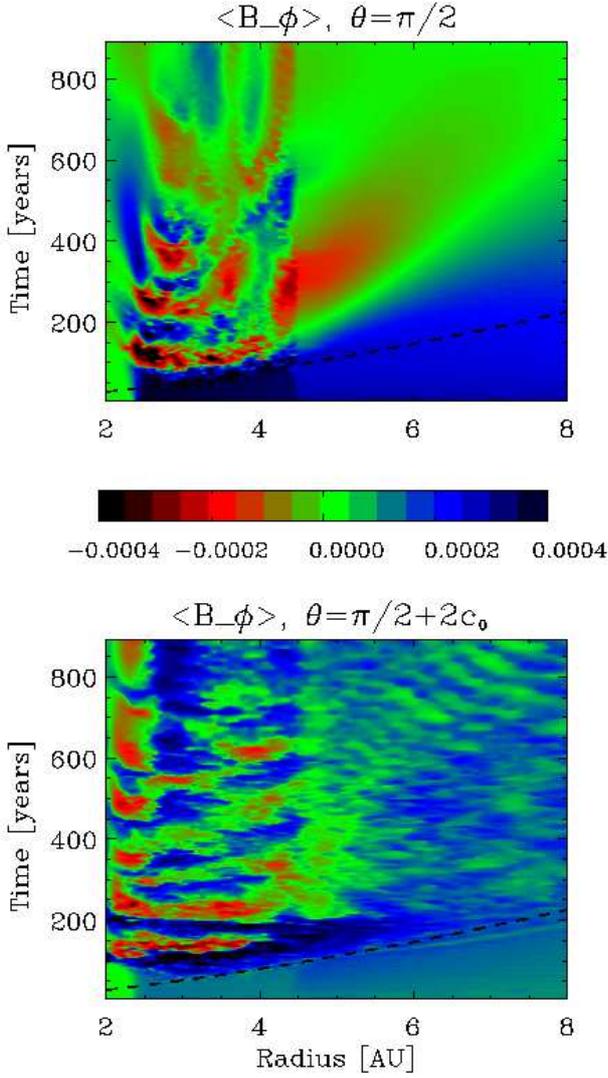}
\caption{ Temporal evolution of azimuthal magnetic field as a function
  of radius for $\rm M_{IR}$ model.  Azimuthally averaged $B_{\phi}$
  field is demonstrated for horizontal slices at
  $\Theta=\pi/2+n\cdot{}c_{0}$ or at $z=n\cdot{}H$, where $n=0,2$.  In
  model $\rm M_{IR}$, the MRI-active zone reaches from $2.5$ to $4.5$
  AU and layer $z=0$ is MRI-`dead' between $4.5$AU and $10$AU.  The
  period of the $B_{\phi}$ sign reversals is about 150 years (30 local
  orbits).  The black dashed line represents the time of 10 local
  orbits for each radius.}
\label{fluxb3}
\end{center}
\end{figure}

Fig.~\ref{Bphi2} demonstrates how the inverse plasma beta,
$1/\beta=P_{\rm mag}/P_{\rm gas}$, changes due to MRI from $P_{\rm
  mag}/P_{\rm gas}=1/25$ (red dotted line) to the convex shape (black
lines).  The solid line stands for inverse plasma beta averaged within
the active zone ($2.5\to 4.5$ AU).  The dashed line stands for the
$1/\beta$ averaged over the patch of the dead zone ($5\to 7$AU).  In
the midplane we find the minimum of magnetic pressure, with $P_{\rm
  mag}/P_{\rm gas}=0.01$.  The plasma beta is reaching 1 at 2.8 H both
in model $\rm M_{IR}$ and in the low-resolution run $\rm M_{IR/2}$
(Fig.~\ref{Bphi2}, left).  The resulting vertical profile of the
magnetic pressure is very similar to those shown in Fromang \& Nelson
(2006).  It is remarkable, that the dead zone builds up the same
vertical distribution of magnetic pressure as the active zone,
predominantly due to the smooth azimuthal magnetic field component.
Radial dependence of the inverse plasma beta (Fig.~\ref{Bphi2}, right)
in the normal resolution run $\rm M_{IR}$ shows that upper layers
possess the constant $P_{\rm mag}/P_{\rm gas}$, whereas the midplane
layers, i.e. from midplane to 2H, are oscillating and slightly
decrease towards the inner radius within the active zone.  The dotted
line in Fig.~\ref{Bphi2} (top right, model $\rm M_{IR}$) shows that
the magnetic pressure falls to zero at $r=5.5\rm AU$ at the
midplane. The reason is a diffusion of the mean azimuthal magnetic
field from the active zone into the dead zone. This diffused field has
the opposite sign to the primordial field. Its time propagation into
the dead zone is shown in Fig.~\ref{fluxb3} (section~3.2).  The
low-resolution model $\rm M_{IR/2}$ shows that the expelling of the
azimuthal field into the corona is not reaching the same extend as in
the normal resolution model for radii $r>6$AU.

This vertical re-distribution of the azimuthal magnetic field affects
the effective resolution.  In the left panels of Fig.~\ref{Bphi2}, we
adopt solid lines for active and dashed lines for dead zone values.
Green vertical bars in Fig.~\ref{Bphi2} mark the midplane region with
$\lambda_{\rm c}/\Delta \phi<8$ (Eq.~\ref{resol}).  The blue vertical
bars show the disk height where Els\"asser number $\Lambda=v_{{\rm
    A}z}^2/\eta\Omega$ drops below unity.  The criterion for MRI
instability $\Lambda> 1$ has been introduced in \citet{san01} for the
case of non-ideal MHD with Ohmic dissipation.  After 300 years, the
$B_{\phi}$ vertical profile is changed so much that the midplane
layers are resolved only with $\lambda_{\rm c}/\Delta \phi \geq 5$,
for example in the active zone (2.5 AU to 4.5 AU) in normal resolution
model (green bars, Fig.~\ref{Bphi2}).  On the other side, model $\rm
M_{IR/2}$ becomes well-resolved in the layers $|z/H|> 2$.  Effective
resolution of $\lambda_c/\Delta\phi >16 $ is reached in the active
layers above the dead zone, where vertical MRI is launched outside of
the $\Lambda=1$ line.  Interesting to note, that the Els\"asser number
$\Lambda=v_{{\rm A}z}^2/\eta\Omega$ drops below unity roughly at the
same height, when we compare normal and low resolution models
(Fig.~\ref{Bphi2}, top left and bottom left).  When looking for
numerical values of $\lambda_{\rm c}/\Delta\phi$ get at the location
of blue bars in the active zone, we find $\lambda_{\rm
  c}/\Delta\phi=6$ for $\rm M_{IR}$ and 4 for $\rm M_{IR/2}$.  These
numbers are only approximate values, because it is difficult to
calculate them accurately at the height at which $\Lambda=1$.  All in
all, there are surprisingly small differences between $\rm M_{IR}$ and
$\rm M_{IR/2}$ models.  The instability occurs in both cases and leads
to similar physics, though the speed of the total field amplification
is slower for $\rm M_{IR/2}$ (Table 1).

Fig.~\ref{mags} demonstrates the Els\"asser number and $\Lambda=1$
criterion in the $(r,\Theta)$ plane of azimuthally averaged disk.  The
solid black line of $\Lambda=1$ marks the locations where the
regeneration of vertical magnetic field cannot be efficient enough.
Models $\rm M_{IR}$ and $\rm M_{IR6}$ have their most active MRI
layers at $2.5$H above the 'dead' midplane.  Comparison of the contour
plots of the Els\"asser number for $\rm M_{IR}$, $\rm M_{IR6}$ and
$\rm M_{IDEAL}$ shows, that MRI-active zones have very similar
appearance.  At the midplane of the active zone, there are
yellow-orange areas of low Elsasser number at $r\sim 4$AU ($\rm
M_{IR}$, $\rm M_{IR6}$), $r\sim 4$AU ($\rm M_{IDEAL}$), and $r\sim
5.3$AU ($\rm M_{IR6}$), which are corresponding to the enhanced gas
pressure.  In section~3.4 we describe the formation of the density
rings at these locations in more details.  Condition
$V_{\phi}/\eta\Omega>10$ is fulfilled in the whole disk in every model
\citep{tur08} and the magnetic fields may be pumped into the dead zone
from the active layers.

\subsection{Oscillating Magnetic Fields and Saturation}

\begin{figure}[ht]
\begin{center}
\includegraphics[width=3.0in]{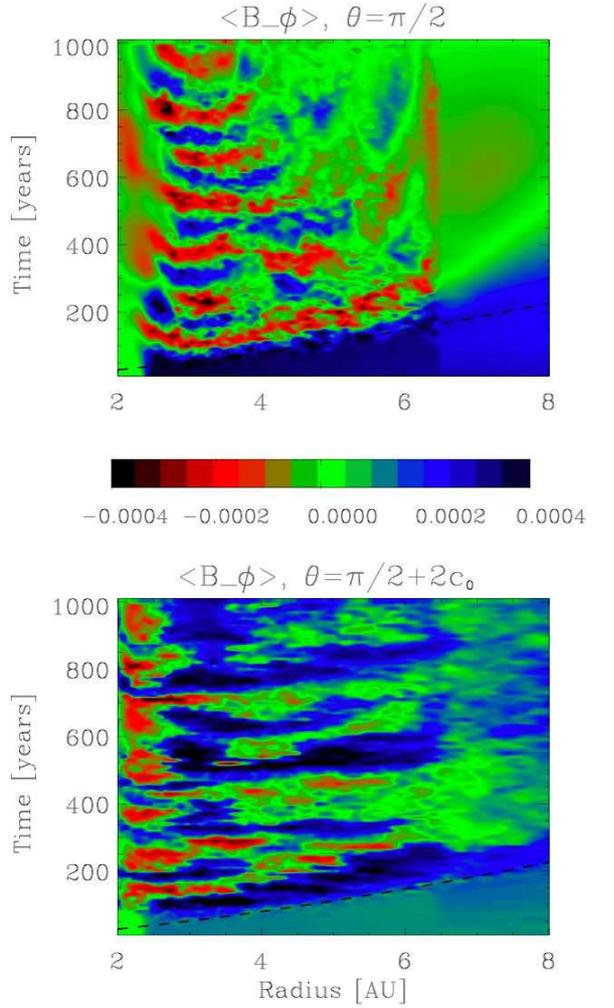}
\caption{ Temporal evolution of azimuthal magnetic field as a function
  of radius for $\rm M_{IR6}$ model.  Azimuthally averaged $B_{\phi}$
  field is demonstrated for horizontal slices at
  $\Theta=\pi/2+n\cdot{}c_{0}$ or at $z=n\cdot{}H$, where $n=0,2$.  In
  model $\rm M_{IR6}$, the MRI-active zone reaches from $2.5$ to $6.5$
  AU.  The period of the $B_{\phi}$ sign reversals is about 150 years
  (30 local orbits), same as in Fig.~\ref{fluxb3}.  Model $\rm
  M_{IR6}$ demonstrates that sign reversal happens not only in time,
  but along the radius as well. The black dashed line represents the
  time of 10 local orbits for each radius.  }
\label{fluxb4}
\end{center}
\end{figure}

\begin{figure*}[ht]
\begin{center}
\includegraphics[width=4.5in]{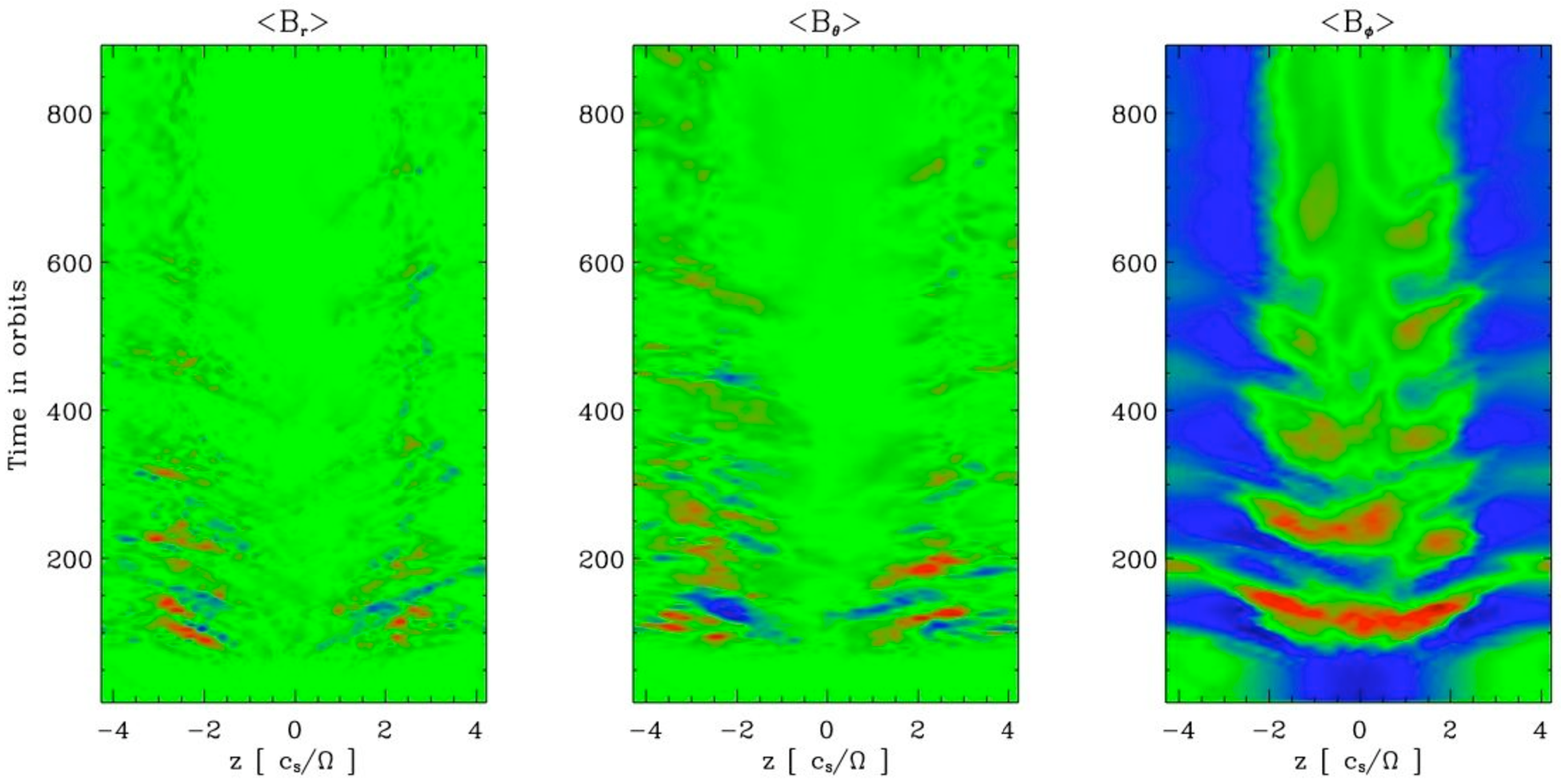} 
\includegraphics[width=4.5in]{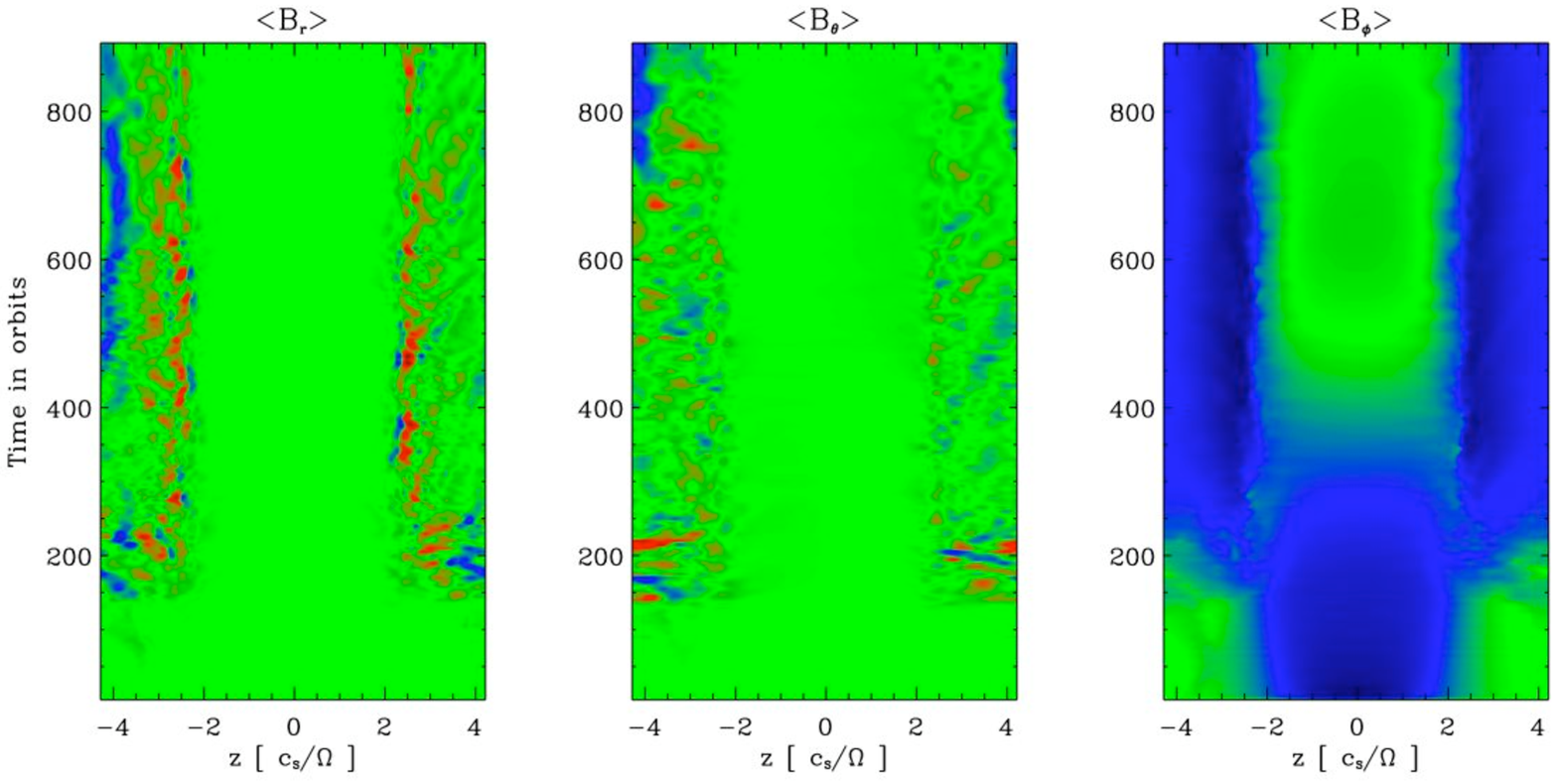}
\caption{Horizontally averaged magnetic fields components
  $\langle{B_{r}\rangle}$, $\langle{B_{\Theta}\rangle}$ and
  $\langle{B_{\phi}\rangle}$ as function of disk height and time. Blue
  and red colors are tracing the positive and negative fields, green
  is always zero.  $\langle{B_{\phi}\rangle}$ sign-reversals `swim'
  from midplane to the disk corona ('butterfly diagram').  Above:
  model $\rm M_{\rm IR}$, averaging of the magnetic fields has been
  done within the MRI-active zone (between 2AU and 4AU).  Below: model
  $\rm M_{\rm IR}$, averaging of the magnetic fields has been done
  within the dead zone (between 6AU and 8AU).}
\label{butter1}
\end{center}
\end{figure*}

After most of the initial azimuthal magnetic flux has been shifted to
the upper disk layers, the two scale heights to the adjacent midplane
develop the oscillating axisymmetric magnetic field.  For model $\rm
M_{IR}$, the oscillations last for over 400 years and then decay.  In
the MRI-active zone, i.e. between 2.5 and 4.5 AU, the $B_{\phi}$
sign-switching occurs within about every 120 to 150 years.  In
Fig.~\ref{fluxb3} we demonstrate the time evolution of the azimuthally
averaged $B_{\phi}$ as a function of the radial distance for the $\rm
M_{IR}$ model. The fluctuating part of the azimuthal magnetic field is
roughly ten times stronger than the mean part, $B_{\phi}^{'} \sim 10
\langle{B_{\phi}\rangle}$. 

\begin{figure}[h]
\begin{center}
\includegraphics[width=3.0in]{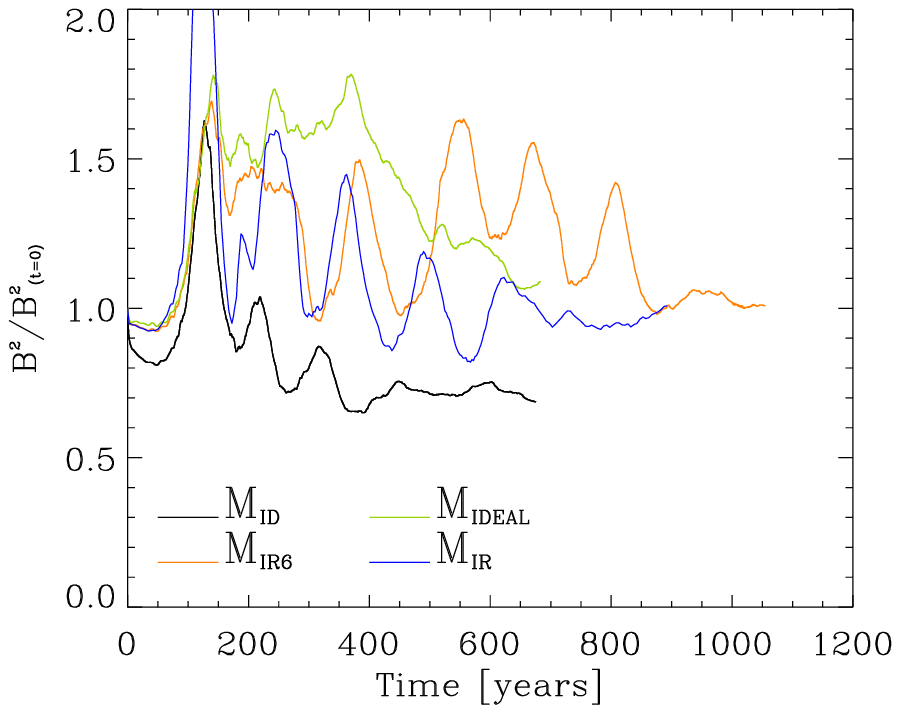} 
\includegraphics[width=3.0in]{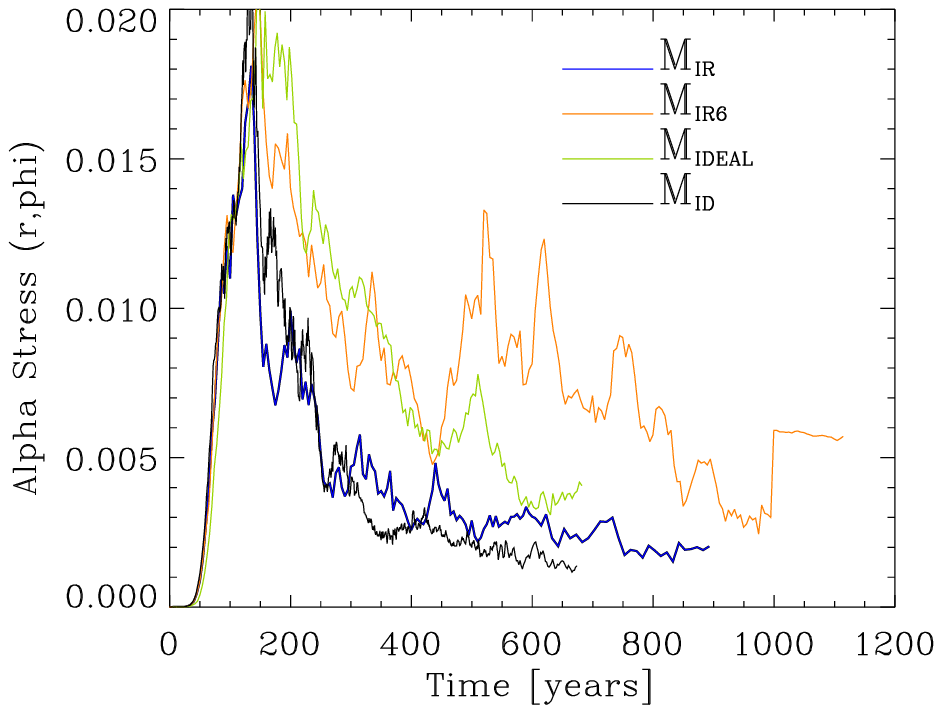} 
\caption{Top: Total magnetic energy evolution; Bottom: Total alpha
  stress evolution for models $\rm M_{IR}$, $\rm M_{ID}$, $\rm
  M_{IR6}$ and $\rm M_{IDEAL}$. The oscillations are correlated with
  the `butterfly diagram'.  }
\label{energ}
\end{center}
\end{figure}

This property appears to be typical for
MRI, as the observations of the magnetic fields in the galactic disks
show (see \citet{bec00} for review).  When the active zone is
stretched up to $6.5$ AU, it is not affecting the time period of the
$B_{\phi}$ sign reversals.  Model $\rm M_{IR6}$ demonstrates that sign
reversals occur not only in time, but also along the radius, as shown
in Fig.~\ref{fluxb4}.  The first positive $B_{\phi}$-stripe is
starting at 60 years at 3 AU and progresses to 6 AU in 160 years
(Fig.~\ref{fluxb4}).  The stripes of $B_{\phi}$ in $(r,t)$-plane are
stretched along the line of local orbit, which is plotted with black
dashed line for $t_{\rm local}=10$ in Fig.~\ref{fluxb4}.  At later
times, the mixing and interaction of the waves is breaking radial
$B_{\phi}$ reversals into less regular oscillations both in
Fig.~\ref{fluxb3} and in Fig.~\ref{fluxb4}.  The field diffusion in
the dead zone outside of $r_{\rm th}=4.5$ AU (model $\rm M_{IR}$) and
outside of $r_{\rm th}=6.5$ AU (model $\rm M_{IR6}$) does not follow
the line of local orbit, $\propto r^{3/2}$, but propagates with
diffusion time $\propto r^{2}$.  For the $\rm M_{IR}$ run, the region
with $r>4.5\ {\rm AU}$ develops the negative azimuthal magnetic fields
due to diffusion of the magnetic field.  After 600 years there is a
weaker wave of positive $B_{\phi}$, which has a shorter time period.

Oscillations in the sign of $B_{\phi}$ at the $z\geq 2$H appear less
clearly, compared to the azimuthal magnetic field at the midplane
(Fig.~\ref{fluxb3}, Fig.~\ref{fluxb4}).  A comparison with $\rm
M_{ID}$ shows why.  The reason is the interaction of MRI waves in the
upper layers between active and dead zones.  For the case of a very
thick dead zone ($\rm M_{ID}$), the sign reversals in $B_{\phi}$ at
$z=2H $ show most pronounced intervals. This is due to the fact, that
MRI can be best exited between 1H and 2H layers of the active zone.
In the model $\rm M_{IR}$, the layers above and below the dead zone
are turbulent and interacting with MRI modes at same height in the
active zone, what leads to a more irregular picture in oscillations of
$B_{\phi}(r,t)$ at $z \geq 2$H.  The color-coded presentation of
$B_{\phi}$ as a function of $(z,t)$ reveals a butterfly diagram, if
the azimuthal magnetic field is averaged within the disk region
between 2 and 4 AU (Fig.~\ref{butter1}, top).  The low plasma $\beta$
does not prevent the upper disk layers from being very turbulent, as
one can see from left and middle panels for $B_{r}$ and $B_{\Theta}$.
Comparing the contours of dominating $B_{\phi}(z,t)$ and other two
turbulent field components $B_{r}(z,t)$ and $B_{\Theta}(z,t)$
demonstrates again that the MRI turbulence and vertical redistribution
of azimuthal magnetic field are connected.  The period of $B_{\phi}$
oscillations is about 150 years, corresponding well to the radial
changes of $B_{\phi}$ sign shown in Fig.~\ref{fluxb3} and
Fig.~\ref{fluxb4}.  When averaging over the whole radial extent of the
disk, or at least within the dead zone, then the butterfly picture
disappears (Fig.~\ref{butter1}, bottom).

\begin{table*}
\begin{center}
  \caption{ $\alpha$-stresses inside (A) and outside (D) of the
    ionization threshold radius for four layers above the
    midplane }
%. The mark $*$ indicates that the steady-state has not yet
%    been reached. 
\label{tab2}
\begin{tabular}{|c|cc|cc|cc|cc|}
\hline
 Models/ &  $\rm M_{IR}$ & & $\rm M_{ID}$ & & $\rm M_{IR6}$ &  & $\rm M_{IDEAL}$ &$*$ \\
 /$\alpha$ stress & (A)  & (D) &  (A)  & (D) & (A)  & (D) & (A)  &  \\
\hline
$\alpha_{\pm{}1H}\cdot{10^{-4}}$ & 2.69 & 0.83 & 2.86 & 1.13 & 4.18 & 1.22 & 5.30&\\
$\alpha_{\pm{}2H}\cdot{10^{-4}}$ & 4.64 & 0.69 & 3.76 & 0.65 & 7.64 & 2.01 & 7.13&\\
$\alpha_{\pm{}3H}\cdot{10^{-4}}$ &10.53 & 7.63 & 7.82 & 0.49 & 15.98 & 11.48& 12.80&\\
$\alpha_{\pm{}4H}\cdot{10^{-4}}$ & 1.14 & 1.00 & 1.75 & 0.59 & 1.41 & 1.62 & 1.93&\\
\hline
\end{tabular}
\end{center}
\end{table*}

Volume-averaging of the magnetic energy shows, that its total value is
oscillating in time, with a period correlated to the sign-switch in
azimuthal magnetic field within $\pm 2H$ relative to the midplane.
Fig.~\ref{energ} shows the magnetic energy for each model in Table 1
and the corresponding total alpha stresses.  The oscillations of
energies in time can be clearly correlated with the butterfly diagram.
Oscillations are weakly visible in the total alpha stress
(Fig.~\ref{energ}, right).  The magnetic energy curves reach a
constant value in model $\rm M_{IR}$, which has the smallest active
zone (from $2.5$ to $4.5$ AU).  Model $\rm M_{ID}$ looses the total
magnetic energy continously during 400 years due to higher magnetic
dissipation.  Model $\rm M_{ID}$ reaches a steady-state when stresses
and magnetic energy remain unchanged from 400 to over 600 years.  The
longer is the MRI-active domain, the longer it takes for the
simulation to reach the steady-state. Models $\rm M_{IR6}$ and $\rm
M_{IDEAL}$ hold oscillatory (non-stationary) magnetic fields for 900
years.

The closed boundary conditions enforce the conservation of the total
flux in the domain.  Fluxes of vertical magnetic field remain zero
through the whole simulation.  The effect of the boundary choice on
the butterfly diagram remains to be investigated in future work. The
local box simulations, made for open vertical boundaries, show the
butterfly picture as well \citep{tur08}.

%%%%%%%%%%%%%%%%%%%%%%%%%%%%%%%%%%%%%%%%%%%%%%%%%%%%%%%%%%%%%%%%%%%%%%%%%

\subsection{Maxwell and Reynolds Stresses}

Radial inhomogeneity in turbulent viscosity has been suggested as the
mechanism to produce the pressure maximum, which is efficient in dust
trapping, and therefor important in the planet formation theory.  In
our models, the turbulent viscosity is driven by MRI and the
inhomogeneity in turbulent stresses appears naturally as the result of
simulations, when we include the sharp gas ionization threshold.
Indeed, we find a density bump forming behind the ionization threshold
in our simulations (section~3.5), and a corresponding jump in
turbulent $\alpha$ stress.

\begin{figure}[h]
\begin{center}
\includegraphics[width=3.0in]{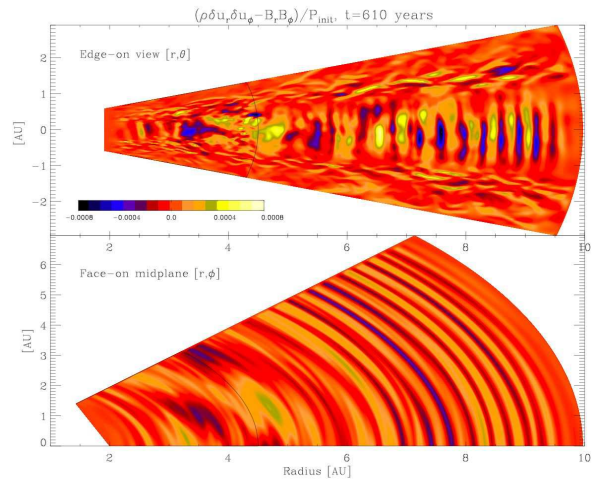}
\caption{Snap-shots of turbulent $\alpha$ stress in $(r,\Theta)$ (top)
  and $(r,\phi)$ (bottom) slices for models $\rm M_{IR}$ for data
  output at $t=610$ years.  Black line indicates the ionization
  threshold.  }
\label{plane}
\end{center}
\end{figure}

\begin{figure*}[ht]
\begin{center}
\includegraphics[width=6.5in]{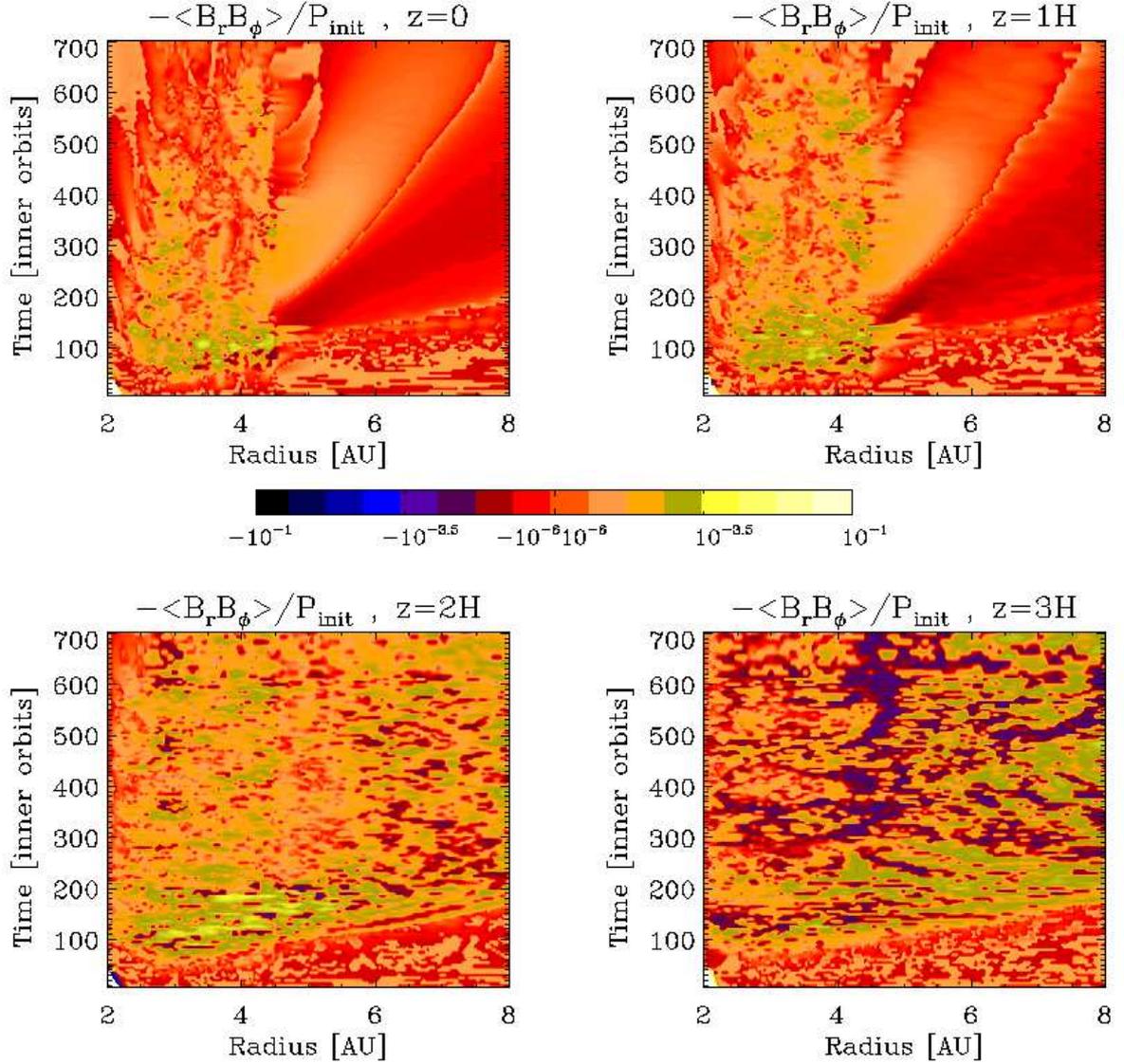}
\caption{Model $\rm M_{IR}$: temporal and radial evolution of Maxwell
  stress. Horizontal slices of $T_{\rm M}(r,t)$ are taken at
  $\Theta=\pi/2+n\cdot{}c_{0}$ or $z=n\cdot{}H$, where $n=0,1,2,3$.
  The color scale shows the azimuthally averaged Maxwell stress in
  local gas pressure units. Dark-orange filaments ($\pm 10^{-7}$)
  in the active zone correspond to the location where the reversals of
  axisymmetric azimuthal field happen.  Weak Maxwell stress of
  $\pm10^{-5} $ also occurs in the dead zone.}
\label{fluxmax}
\end{center}
\end{figure*}

The time evolution of the Maxwell stress $T_{\rm M}(r,t)$
(Eq.~\ref{10}) is demonstrated in Fig.~\ref{fluxmax}.  There is a weak
Maxwell stress of about $10^{-5}$ in the dead zone, which can
periodically become negative.  One can see the sharp border in $T_{\rm
  M}(r,t)$ between the active and the dead zones in slices for $z=0$,
$z=1H$ and $z=2H$.  The traces of sign reversals in the azimuthal
magnetic field are also visible in the Maxwell stress.
Fig.~\ref{fluxmax} shows that exactly at 150 years the Maxwell stress
reaches its maximum of $10^{-1}$, when calculated in units of initial
pressure $\langle{P(r)\rangle}$ (Eq.~\ref{10}).  Later on, the
saturation of MRI sets in and the total stress is between $10^{-3}$
and $10^{-2}$ (see Table 1).  The dark-orange filaments of very weak
Maxwell stress in the active zone, $\pm 10^{-7}$, correspond to the
location where the reversals of axisymmetric azimuthal field happen.
The weak Maxwell stress is located at $r=3.5$ AU for many years, what
appears as a systematic stripe when looking at $z=0$ and $z=1H$
horizontal slices of $T_{\rm M}(r,t)$ in Fig.~\ref{fluxmax}.  This is
a location where the density ring is created (more in section~3.4) and
also most of $B_{\phi}$ reversals take place during the time period
from 200 to 700 years.  Negative values of Maxwell stress appear at 3H
above the midplane. This is the region of low plasma beta, and the
turbulence at this height is no more MRI-driven.

In the dead zone, the value $\alpha_{\rm total}\sim{10^{-3}}$ is due
to Reynolds stress contribution.  In order to provide the
understanding of how the turbulence there looks like, we present a
snap-shot of turbulent $\alpha$ stress for the $\rm M_{IR}$ model
(Fig.~\ref{plane}).  The dead zone is filled with vertical pillars of
the $\alpha$ stress of opposite sign.  In the $(r,\phi)$ plane, those
pillars look like tightly-wrapped spirals.  The spiral waves are
launched from the dead-zone edge ($r_{\rm th}=4.5$ AU), where the
non-axisymmetric fluctuations in all velocity components are
MRI-generated.  The weak spiral structures can be found in the gas
density as well.

\begin{figure*}[th]
\begin{center}
\hbox{
\includegraphics[width=3.2in]{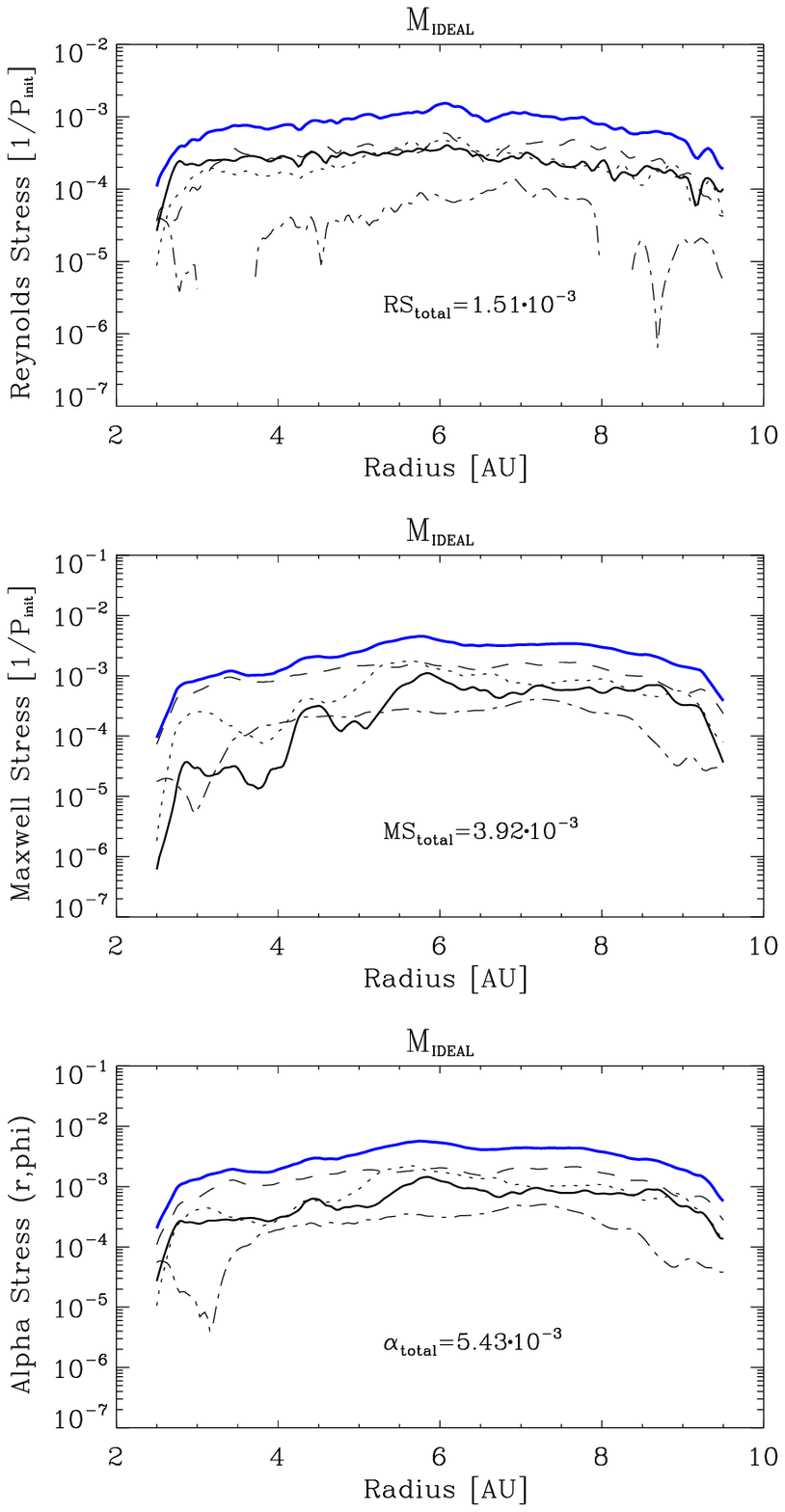}
\includegraphics[width=3.2in]{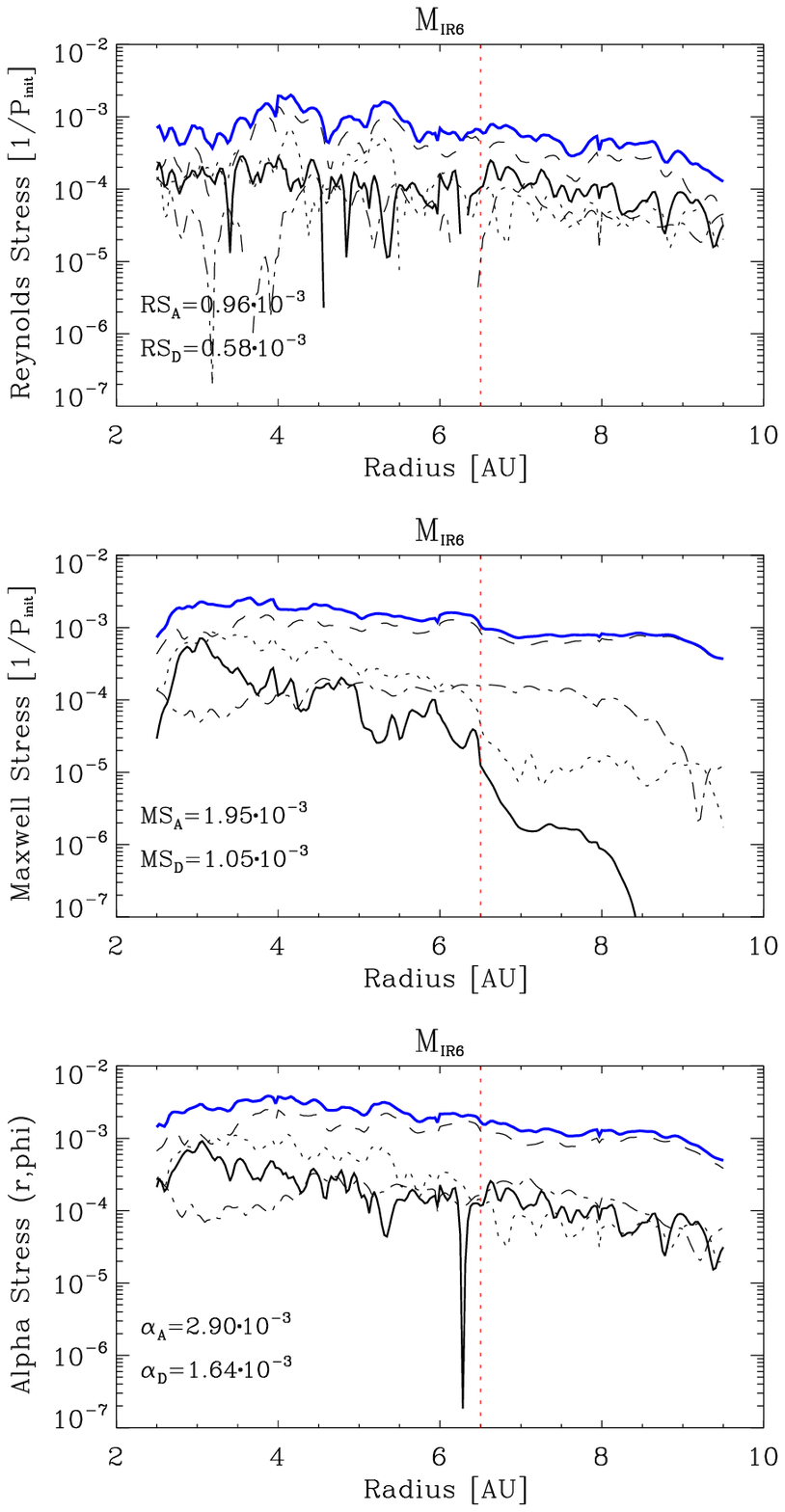}
}
\caption{Reynolds, Maxwell and total $\alpha$ stresses for models $\rm
  M_{IDEAL}$ (left) and $\rm M_{IR6}$ (right).  Vertical averaging
 of stresses is made separately for four midplane-symmetric layers,
  $\pm{}\ 1H({\rm solid\ line}), \ 2H({\rm dotted\ line}), \ 3H({\rm
    dashed\ line}), \ 4H({\rm dot-dashed\ line})$. Solid blue line
  shows the classical total stress. All stresses are normalized to the
  pressure at $t=0$.  Time average is made between 850 and 1050 years
  for $\rm M_{IR6}$, and between 610 and 681 years for $\rm
  M_{IDEAL}$.  }
\label{alp2}
\end{center}
\end{figure*}

\begin{figure*}[ht]
\begin{center}
\hbox{
\includegraphics[width=3.2in]{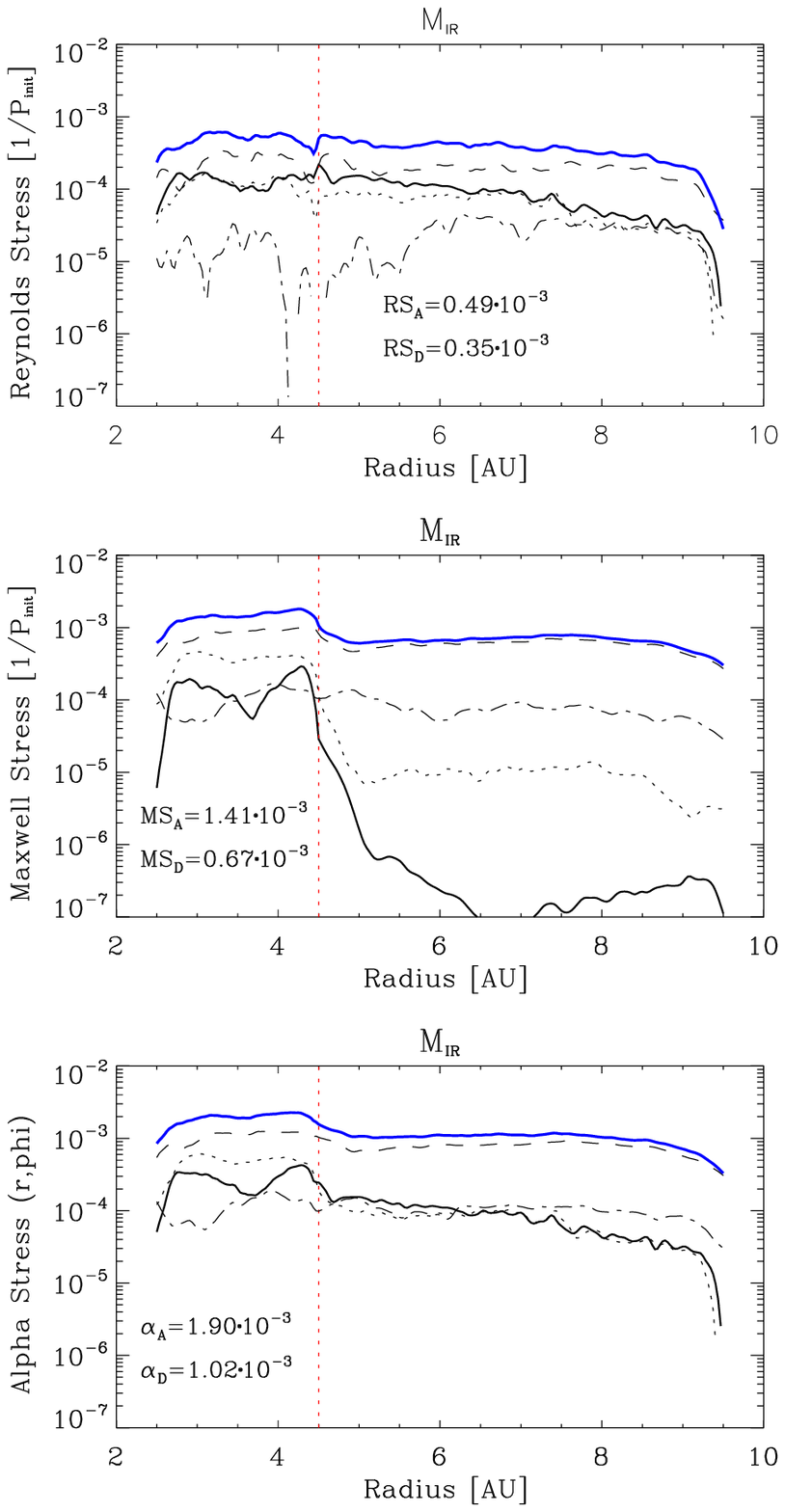}
\includegraphics[width=3.2in]{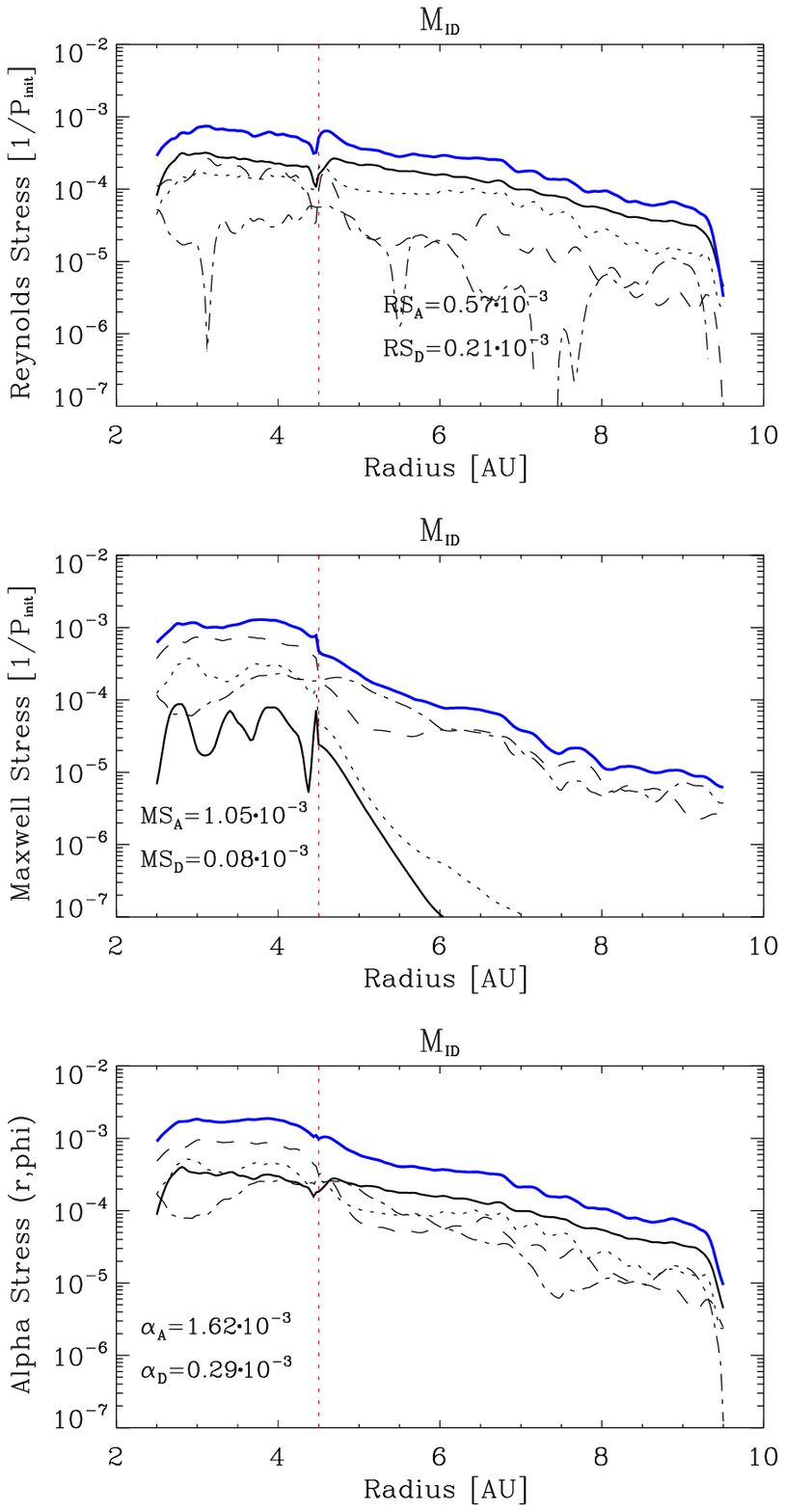}
}
\caption{Reynolds, Maxwell and total $\alpha$ stresses for models $\rm
  M_{IR}$ (left) and $\rm M_{ID}$ (right). Vertical averaging
  of stresses is made separately for four midplane-symmetric layers,
  $\pm{}\ 1H({\rm solid\ line}), \ 2H({\rm dotted\ line}), \ 3H({\rm
    dashed\ line}), \ 4H({\rm dot-dashed\ line})$. Solid blue line
  shows the classical total stress. All stresses are normalized to the
  pressure at $t=0$. Time averages has been done starting from time of
  400 years until the last data output.  }
\label{alp1}
\end{center}
\end{figure*}

We have calculated the turbulent stresses as the function of radius.
The vertical averaging has been done separately for four
midplane-symmetric layers: $|\Theta-\pi/2|<c_{0}$,
$c_{0}<|\Theta-\pi/2|<2c_{0}$, $2c_{0}<|\Theta-\pi/2|<3c_{0}$ and
$3c_{0}<|\Theta-\pi/2|<4c_{0}$, where $c_{0}=H/R=0.07$.  The resulting
$\alpha$ stresses for these layers are given in Table~\ref{tab2} as
$\pm{}1H,\ 2H,\ 3H,\ 4H$ correspondingly and plotted in
Fig.~\ref{alp2} and Fig.~\ref{alp1} with solid, dotted, dashed and
dot-dashed black lines. 
 In Table\ref{tab2}, the mark $*$ indicates that the steady-state has not yet
   been reached.
 Solid blue lines in Fig.~\ref{alp2} and
Fig.~\ref{alp1} represent the $\alpha(r)$ integrated over the whole
disk thickness.  The strongest total stress alpha of all models is
obtained in model $\rm M_{IDEAL}$, which remains of the same order of
magnitude with radius.

The turbulent stresses have $\propto r^{-2}$ slope during the
'butterfly' evolution stage in models $\rm M_{IR6}$ and $\rm M_{IR}$.
For the $\rm M_{IR}$ model, the time averaging of the stresses between
250 and 600 years results in a $\alpha\propto r^{-2}$.  Afterwards,
the turbulence in $\rm M_{IR}$ reaches a 'butterfly'-free state
($t>600$ years). When the temporal averaging is made between 400 and
800 years, instead of from 250 to 600 years, we obtain $\alpha_{\rm
  A}$ which is constant with radius (Fig.~\ref{alp1}).
In the $\rm M_{IR}$ model, the Maxwell stress is the main contribution
to $\alpha_{\rm total}$ only for layers $\ 3H,\ 4H$ outside the
threshold $r_{\rm th}=4.5\rm AU $ (Fig.~\ref{alp1}).

The gas density perturbations look like spiral waves, and the Maxwell
stress is strongly reduced outside of the ionization threshold.  Model
$\rm M_{ID}$ shows a steeper fall in Reynolds stress and in total
$\alpha$ in the dead zone, compared to the $M_{\rm IR}$ model.  The
active layers above and below the dead zone are very thin and only
marginally unstable to MRI. Model $\rm M_{ID}$ has $1H,\ 2H$ and
partly $3H$ layers which are deactivated in the dead zone.  The
pumping of the waves happens mostly from the MRI-active zone, and not
from both active zone and adjacent layers as in $\rm M_{IR}$.
Comparing models $\rm M_{IR}$ and $\rm M_{ID}$, we conclude that the
pumping of the hydro-dynamical waves into the dead zone is coming both
from 'sandwich' active layers and from the active zone within $r_{\rm th}$.

The summary of turbulent $\alpha$ stress values for each layer above
the midplane is presented in Table~\ref{tab2}. The radial averaging
has been done separately for active (A) and dead (D) zones. The
turbulent $\alpha_{\rm(A)}$ stress is increasing from midplane to
$2c_{0}<|\Theta-\pi/2|<3c_{0}$, and drops in the fourth layer. The
most prominent decrease in $\alpha$ stress from active to dead zone
happens within adjacent to the midplane $|\Theta-\pi/2|<2c_{0}$ layers
for models $\rm M_{IR}$ and $\rm M_{IR6}$. In the case of very thick
dead zone ($\rm M_{ID}$), the decrease in turbulent stress at the dead
zone edge is significant in all three layers, $|\Theta-\pi/2|<3c_{0}$.
%%%%%%%%%%%%%%%%%%%%%%%%%%%%%%%%%%%%%%%%%%%%%%%%%%%%%%%%%%%%%%%%%%%%%%%%%

\subsection{Development of Pressure Maxima and Trapping of Solids }

Due to active secretion through MRI-active zone, the pressure bump at
the inner edge of the dead zone is formed within a hundred of inner
orbits.

\begin{figure*}[ht]
\begin{center}
\hbox{
\includegraphics[width=3.in]{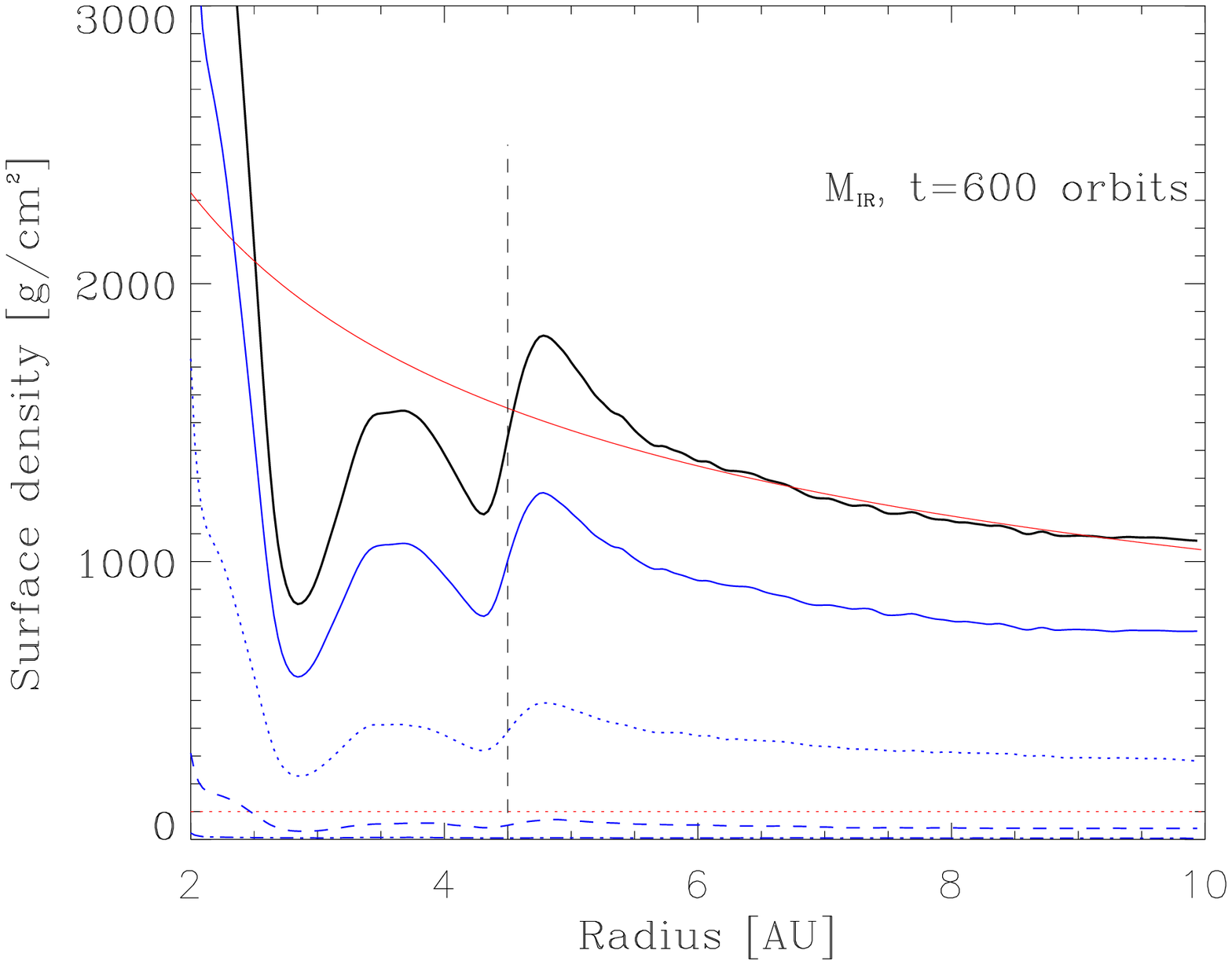}
\includegraphics[width=3.in]{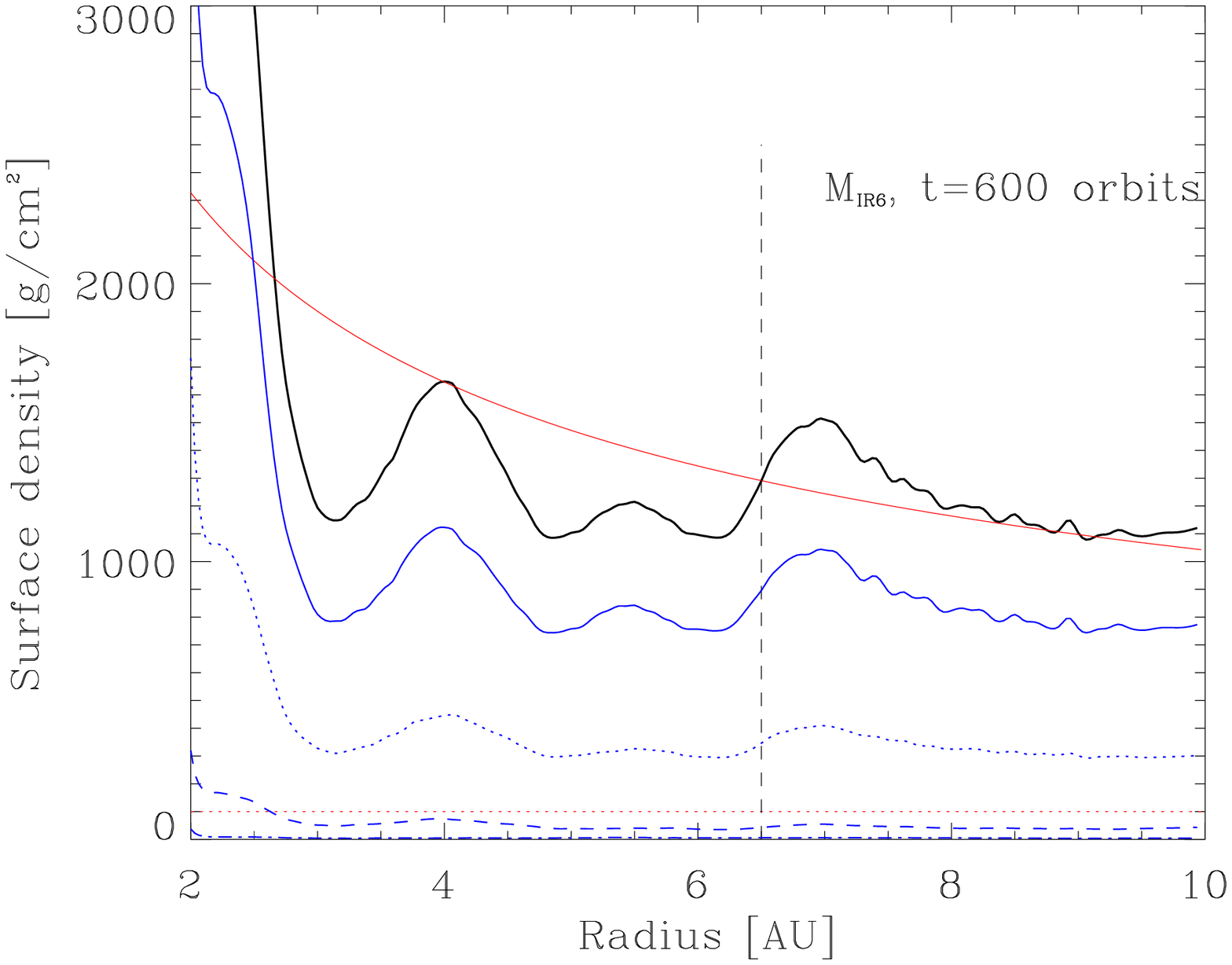}
}
\hbox{
\includegraphics[width=3.in]{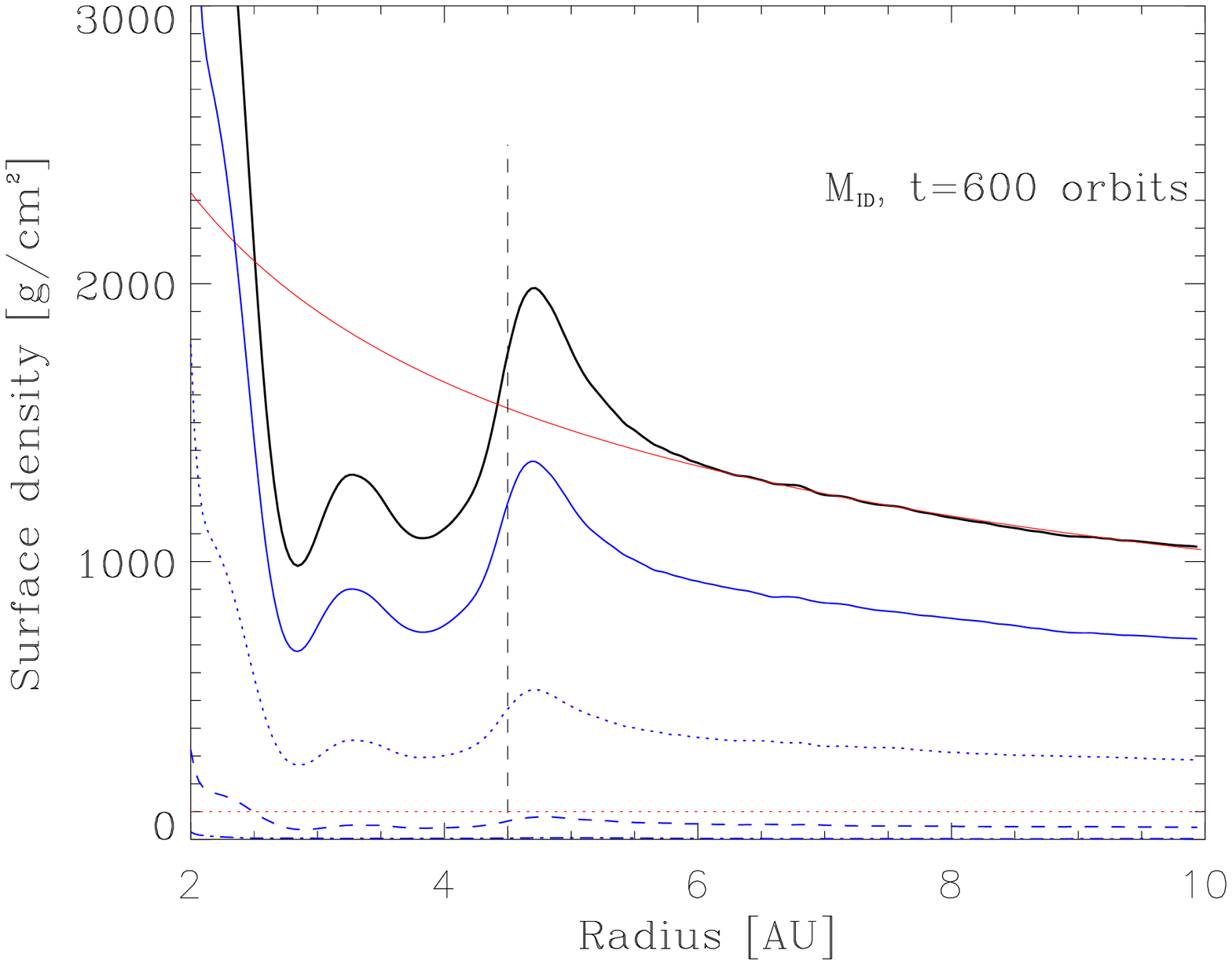}
\includegraphics[width=3.in]{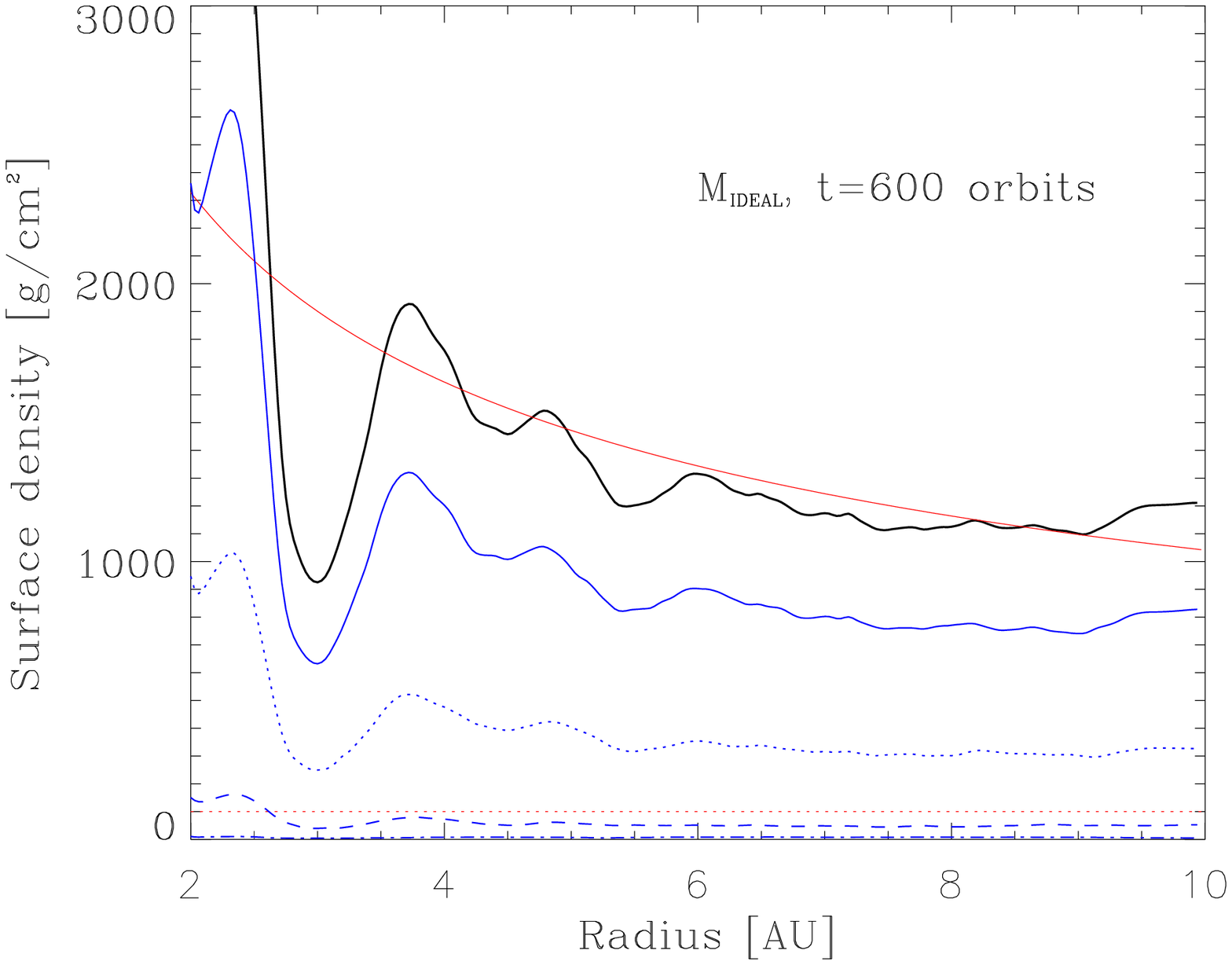}
}
\caption{Surface density after 600 years of evolutions.  Red solid
  line stands for the initial surface density profile,  black solid
  line represents the final surface density. Red dotted line indicates
  the cosmic ray adsorbtion depth $100\rm g/cm^2$. Blue lines show the
  contributions of every scale height in the surface density.
  Vertical dashed line shows the inner edge of the dead zone.}
\label{surd1}
\end{center}
\end{figure*}

The change in the radial density gradient appears in the disk layers
starting from the midplane and up to 3H.  The piling up of the density
behind the $r_{\rm th}$ is accompanied by a broad gap before the
threshold. In addition, we find the rings of enhanced density within
the MRI-active zone.  The pressure bump at the inner edge of the dead
zone and the rings of enhanced density within the MRI-active
zone are formed due to different processes, which we discuss in
section~4.  Fig.~\ref{surd1} (top) shows the changes of surface
density for models $\rm M_{IR}$ and $\rm M_{IR6}$.  In our
simulations, the number of density rings depends on the extension of
the active zone.  In the model without a dead zone, there are three
rings of enhanced density appearing within the domain before the quasi
steady-state is reached (Fig.~\ref{surd1}, bottom right).  The thicker
dead zone (model $M_{\rm ID}$) seems to be more efficient in piling up
the higher bump: the maximum of the surface density peak is larger,
when compared to the snap-shots of surface density in $\rm M_{IR}$ at
the same time.

\begin{figure*}[ht]
\begin{center}
\hbox{
\includegraphics[width=4in]{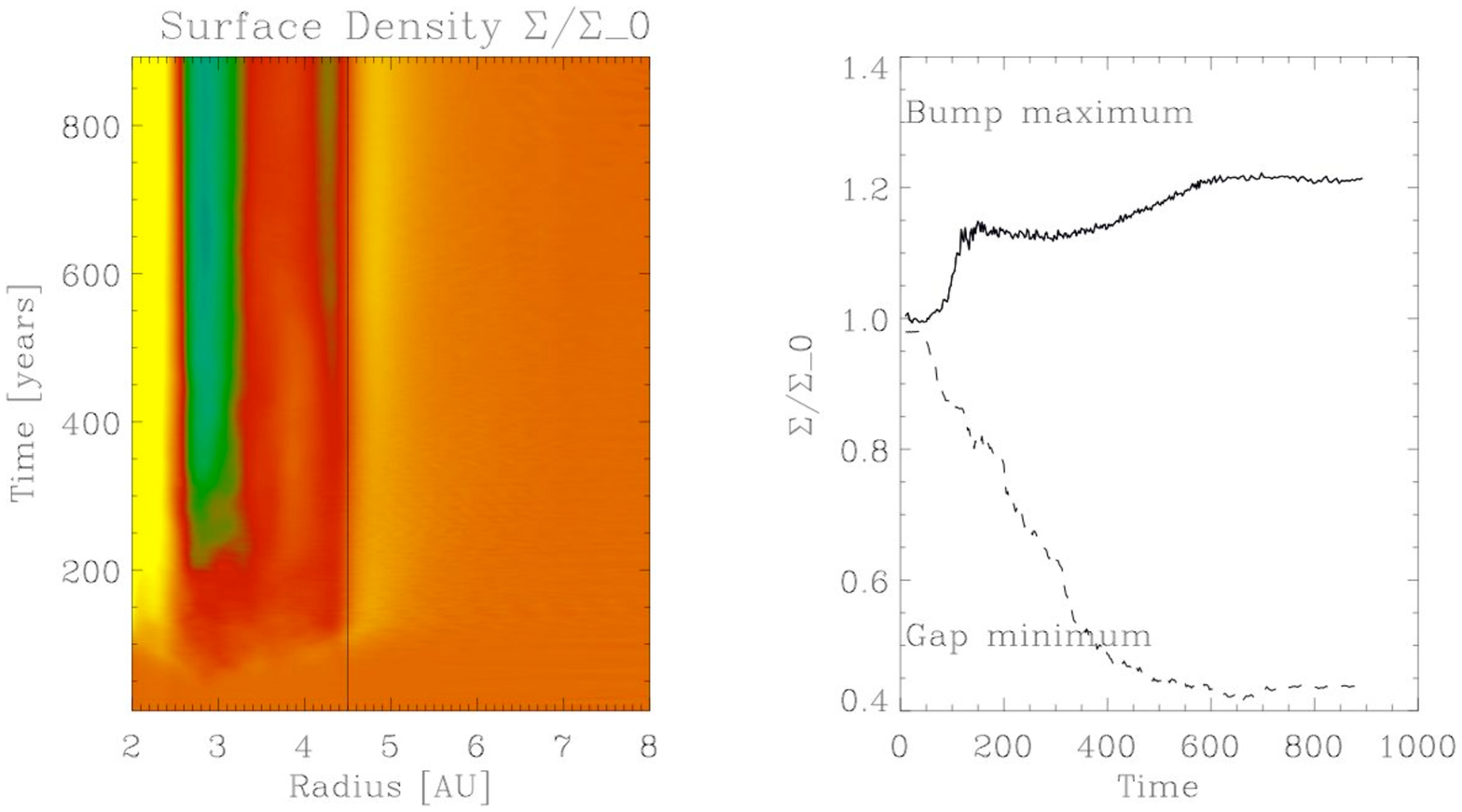}
\includegraphics[width=2.18in]{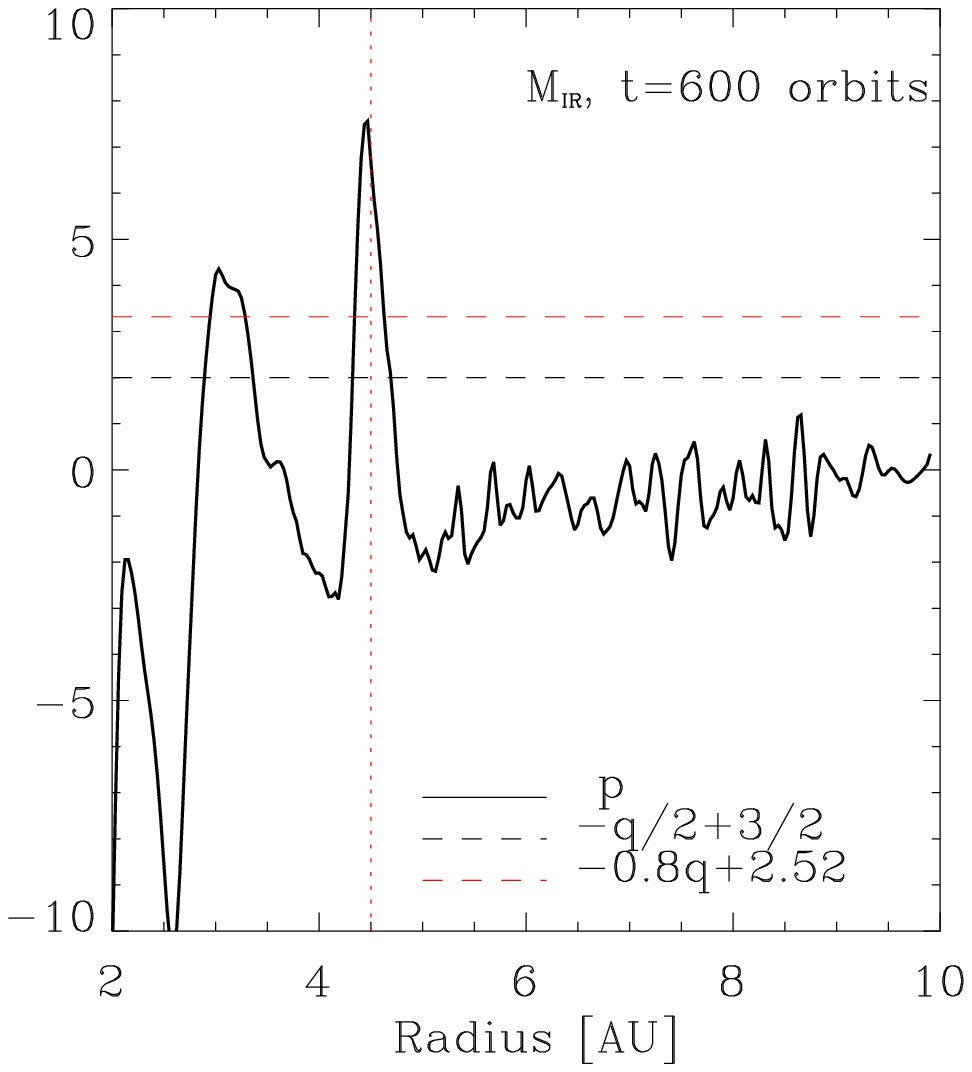}
}
\hbox{
\includegraphics[width=4in]{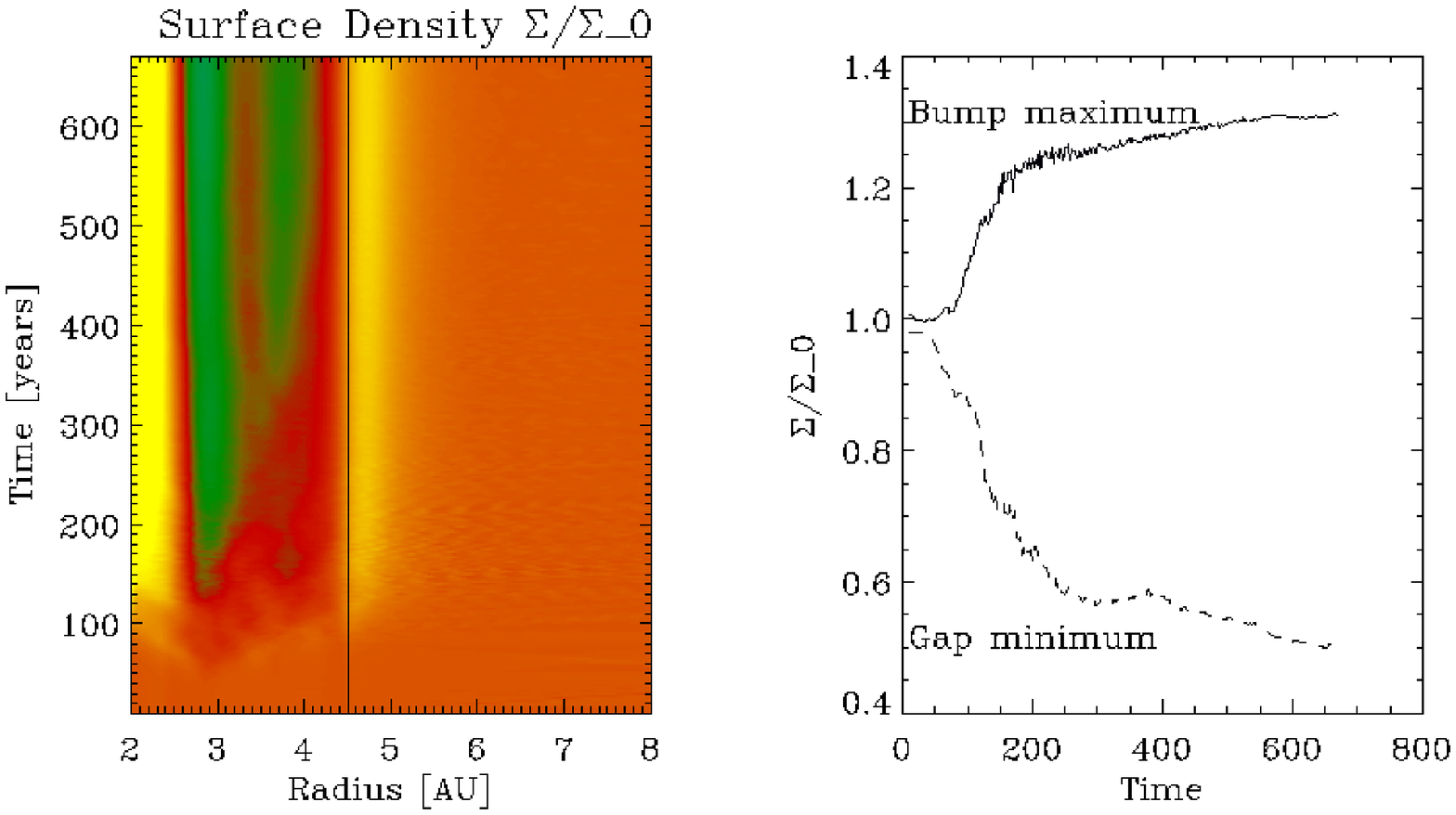}
\includegraphics[width=2.18in]{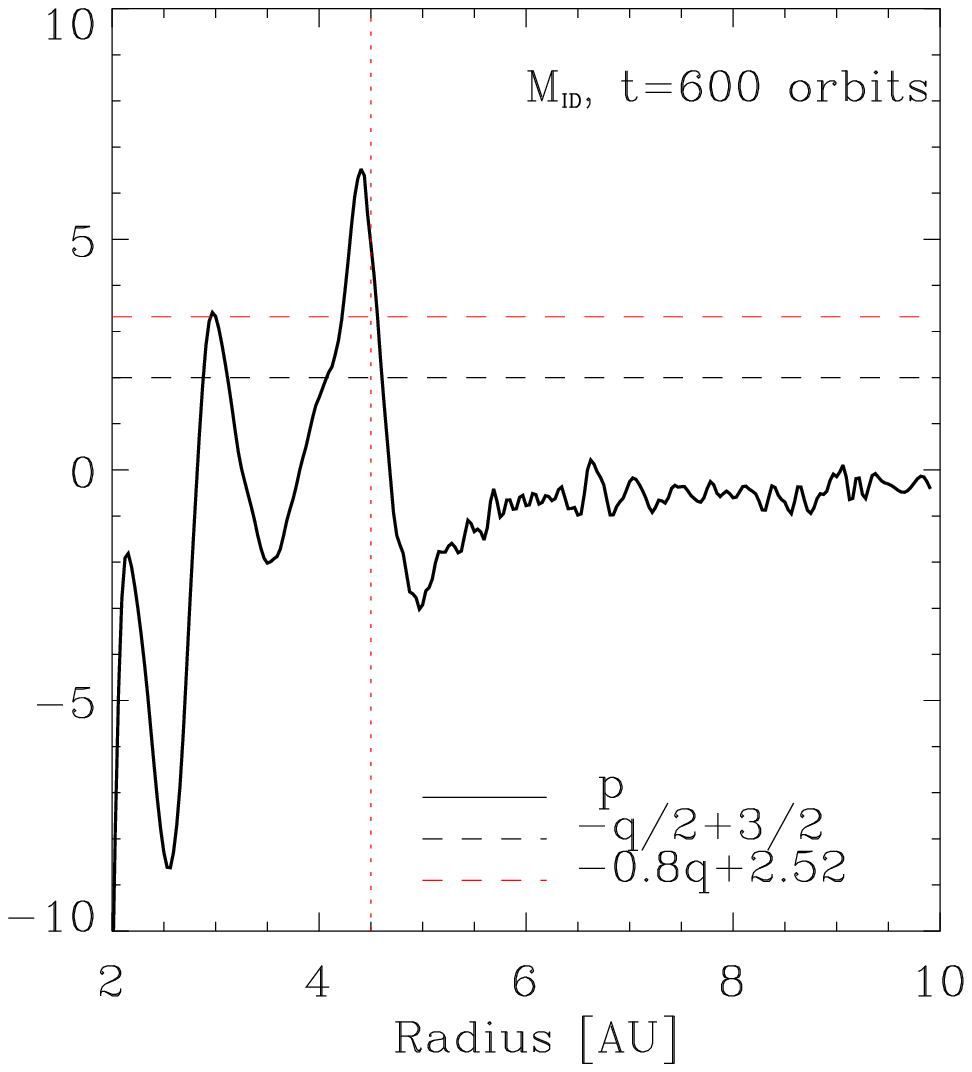}
}
\hbox{
\includegraphics[width=4in]{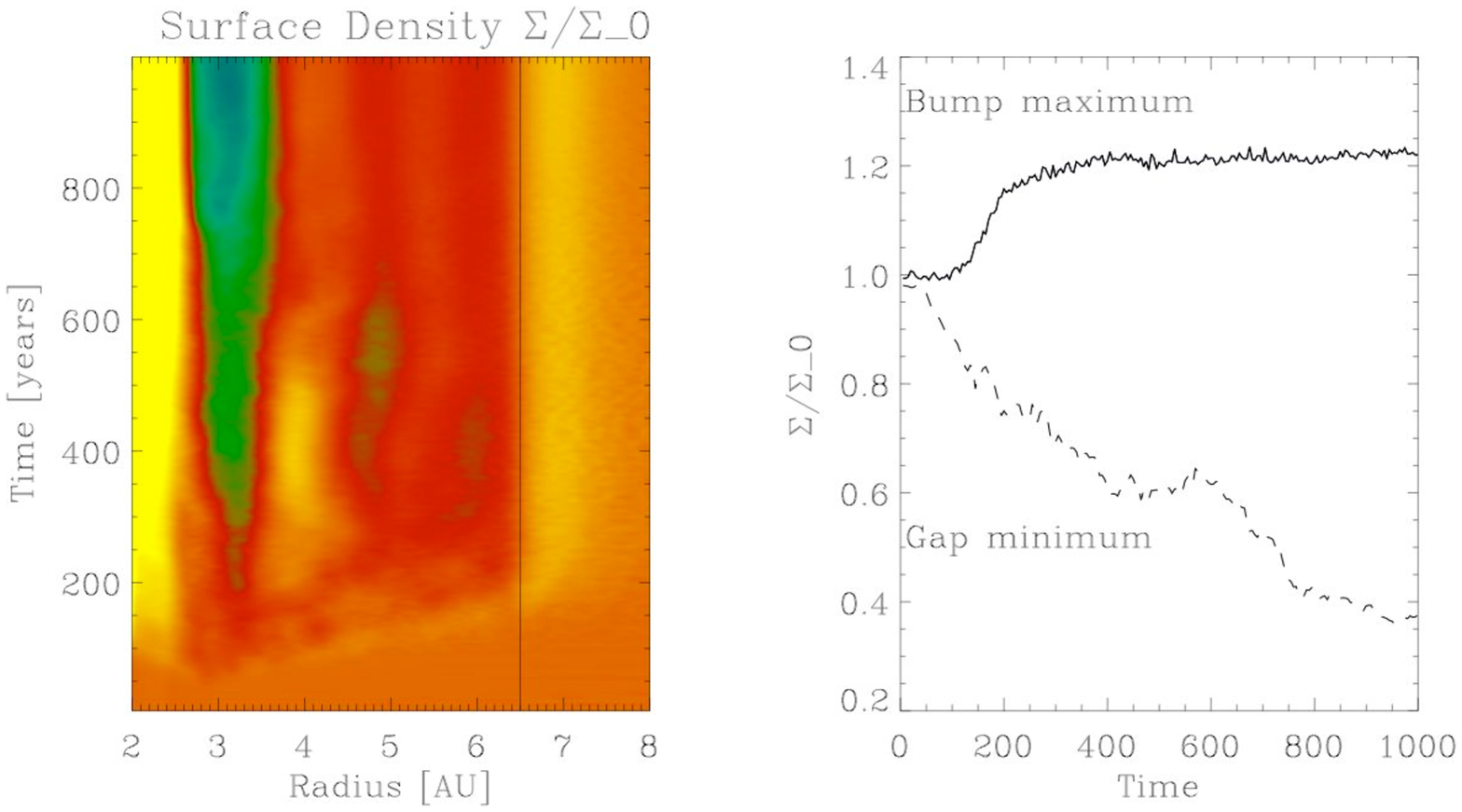} 
\includegraphics[width=2.18in]{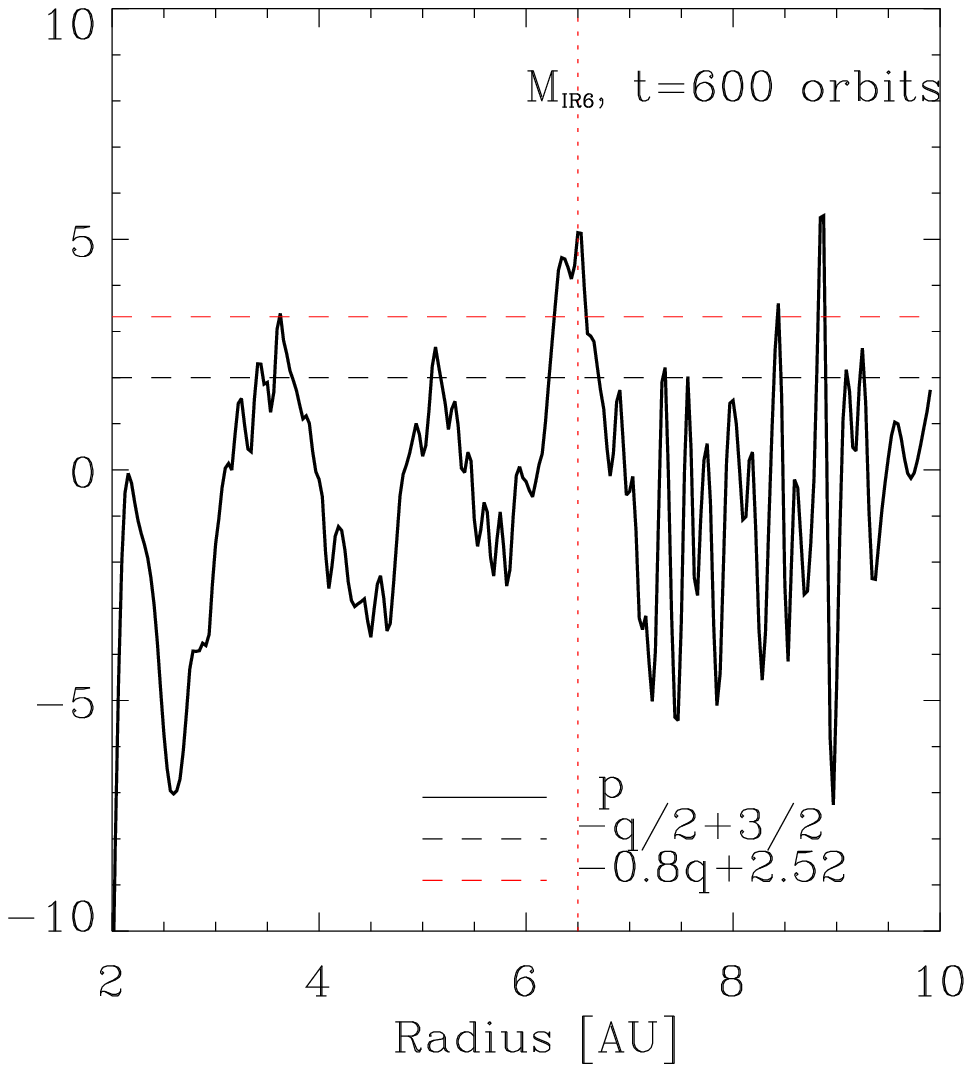}
}
\caption{Relative surface density $\Sigma/\Sigma_{0}$, for models
  $M_{\rm IR}$ (above), $M_{\rm ID}$ (middle) and $M_{\rm IR6}$
  (below).  Left: Colors show the evolution of density bumps
  (yellow) and  gaps (green).  Middle: time development of
  $\max(\Sigma/\Sigma_{0} )$ of the bump at $r=[3:6] $AU and
  $\min(\Sigma/\Sigma_{0} )$ of the gap in $r=[2.5:5]$AU.  Right:
  Criteria for outward migration (Eqs.~\ref{Ida1},~\ref{Ida2}).  Rings
  of enhanced density are formed in the active zone and the outmost
  density ring is located at the ionization threshold, marked as a
  vertical line in both left and  right panels.}
\label{dens}
\end{center}
\end{figure*}

First, we consider the accumulation of the density at the inner edge
of the dead zone.  The pressure bump is fixed in time at the location
behind the ionization threshold $r_{\rm th}$, as in Fig.~\ref{dens}
(left).  There are three stages in the evolution of the pressure
maximum at the inner edge of the dead zone (Fig.~\ref{dens}, middle),
for example when considering model $\rm M_{IR}$. One can recognize the
period of very fast mass excavation for the time from $t=0$ to $150 $
years.  During $t=150 \to 600$ years the peak of surface density is
still growing at a roughly ten times slower speed.  After $t=600$
years there is no further increase of surface density, what
corresponds to the steady-state in the presence of saturated
MRI-turbulence.  The most straight-forward explanation for three
stages in density excavation gives a time-dependent evolution of the
Maxwell stress, since it governs the accretion in the MRI-active zone
and in the active layers.  The quasi-steady state is reached for
models $\rm M_{IR}$ and $\rm M_{ID}$ after 600 years, and the maximum
of the pressure bump at the ionization threshold and the minimum of
the surface density in the gap remain unchanged.  The density rings in
the active zone appear to be long-living but less stable features.  We
observe the merging of two rings of enhanced density after 640 years
in model $\rm M_{IR6}$ (Fig.~\ref{dens}, bottom).

The radial pressure gradient is negative in the smooth unperturbed
disk, what leads to sub-Keplerian gas rotation.  Dust grains undergo
an orbital decay \citep{ada76}, because they experience the 'head'
wind. For example, the meter-size particle will migrate from 1 AU into
the Sun within few hundred years (MMSN model).  When local positive
exponent of the disk midplane pressure appears, the dust grains may
experience 'tail' wind and the hydrodynamical drag will lead to their
outward migration \citep{nak86}.  The criterion for outward migration
of the dust grain due to the gas drag is
\begin{equation}
p>-q/2+3/2,
\label{Ida1}
\end{equation}
where $p=d \log{\Sigma_{\rm gas}}/d \log{r}$ and $q=2 d \log{c_{\rm
    s}}/d \log{r}$.  Note that the factor $3/2$ is the normalized
shear.  The planetary embryos can be stopped in such pressure traps as
well \citep{zha08}.  Criterion for outward migration of the
protoplanetary cores has been given in \citet{ida08a} and is based on
the migration rate of planets \citep{tan02}.  When curvature effects
on the Lindblad resonances have been included, the condition for
outward migration for embryos is
\begin{equation}
p>-0.80 q+ 2.52.
\label{Ida2}
\end{equation}
The right panels in Fig.~\ref{dens} show where criteria for outward
migration are fulfilled in our models.  Black lines stand for
criterion given in Eq.~\ref{Ida1} and the red dashed line is for
Eq.~\ref{Ida2}.  The outward migration of planetesimals is
possible within the pressure bump at the inner edge of the dead zone
(Fig.~\ref{dens}).  Conditions for embryos are more difficult to
satisfy.  Not all density rings within the active zone provide
sufficiently strong positive pressure gradients, so that planetary
embryos cannot be stopped there.  The exception is model $\rm
M_{IR}$. There we obtain two pressure traps at $r=3$ AU and at
$r=4.5$AU, where large bodies can be retained.

\subsection{Super-Keplerian Rotation and Similarity to Zonal Flows}

\begin{figure*}[ht]
\begin{center}
\includegraphics[width=6.4in]{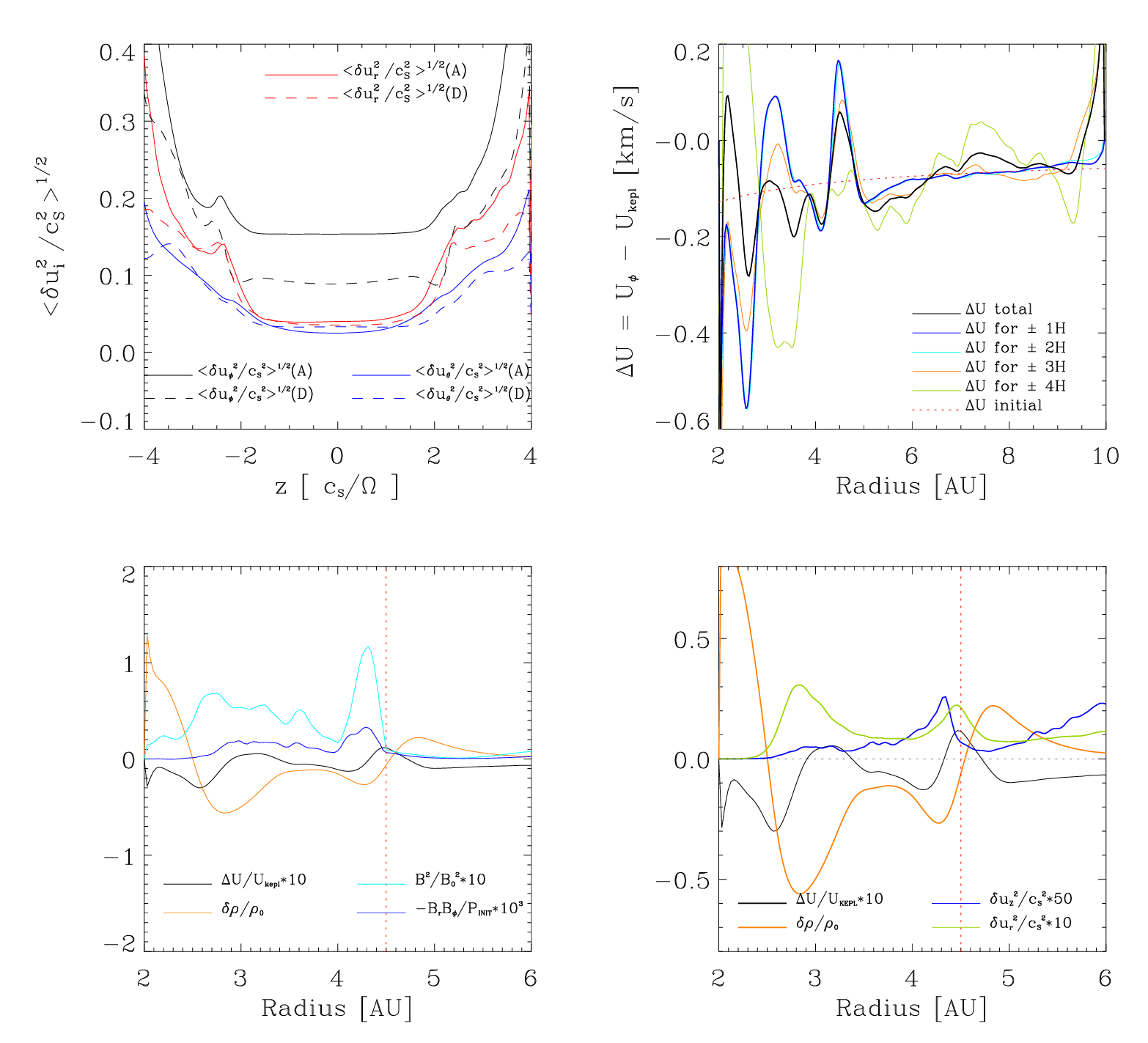}
\caption{Radial and vertical turbulent properties in model $\rm
  M_{IR}$.  Top left: root-mean-square turbulent velocities are
  averaged separately for active (A) and for dead (D) zones.  Top
  right: Outward migration velocity.  Panels on bottom right and
  bottom left show  time-averaged density, Maxwell stress, magnetic
  energy and r.m.s. velocities at the midplane.  Time average is from
  600 to 800 years (steady-state stage in model $\rm M_{IR}$).}
\label{rot1}
\end{center}
\end{figure*}

\begin{figure*}[ht]
\begin{center}
\includegraphics[width=6.4in]{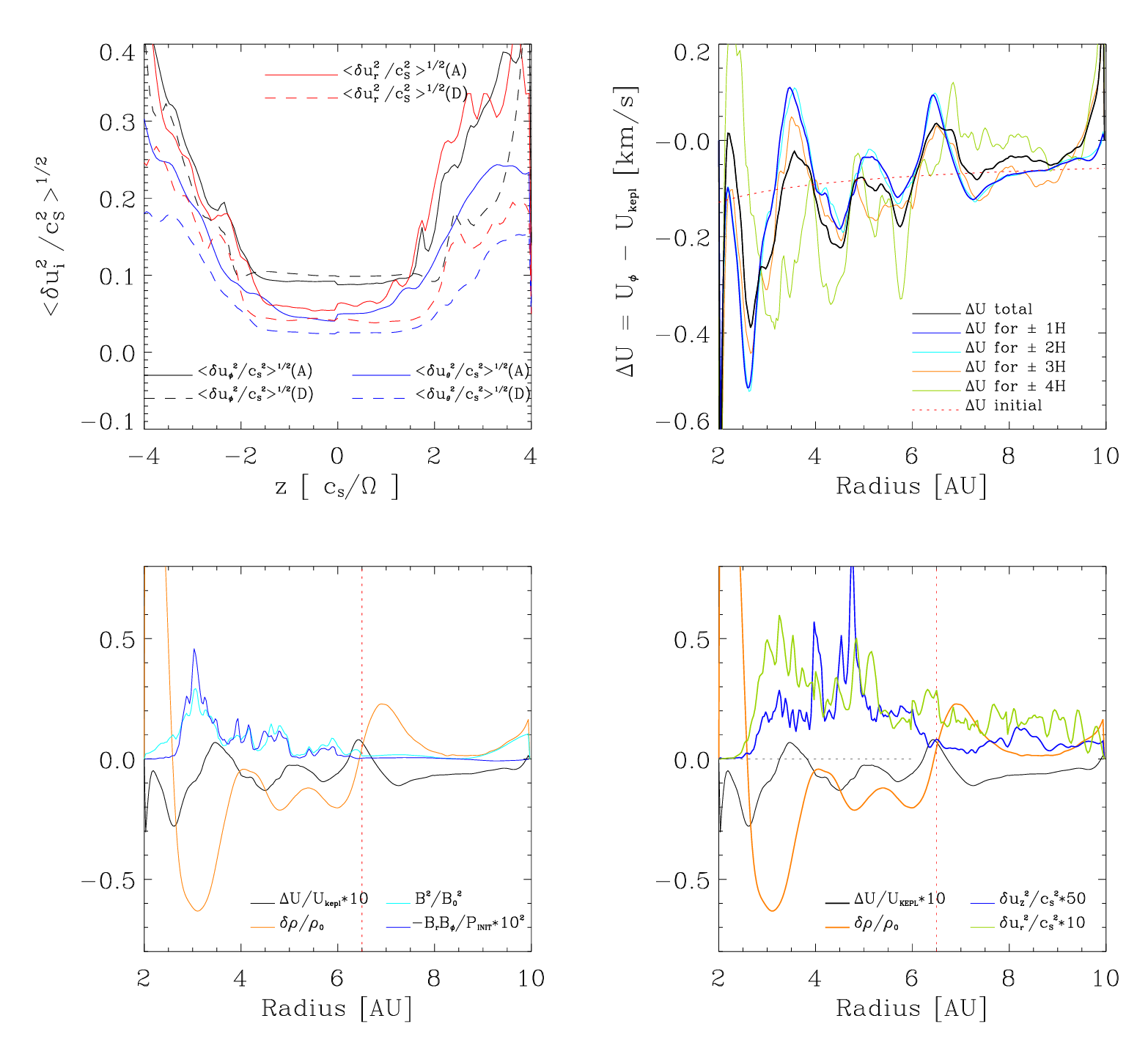}
\caption{Radial and vertical turbulent properties in model $\rm M_{IR6}$.
  Top left: root-mean-square turbulent velocities are averaged
  separately for active (A) and for dead (D) zones.  Top right:
  Outward migration velocity.  Panels on bottom right and bottom left
  show time-averaged density, Maxwell stress, magnetic energy and
  r.m.s. velocities at the midplane.  Time average is from 900 to 1100
  years.
}
\label{rot2}
\end{center}
\end{figure*}

When a quasi-steady state is reached, all time derivatives can be
neglected and the Navier-Stokes equation for radial velocity gives:
\begin{equation}
-\frac{u_{\phi}^2}{r}=\frac{\partial \Psi}{\partial r} - 
 \frac{1}{\rho}\frac{\partial c_{\rm s}^2 \rho }{\partial r}+F_{\rm lorentz}.
\end{equation}
In our locally-isothermal simulations, the temperature is constant on
cylinders.  The steady-state solution is then given as
\begin{equation}
-\frac{(u_{\phi}^2-u_{\rm kep}^2)}{u_{\rm kep}^2} = -\frac{r c_{a}^2}{\sin{\Theta}}
             \left(
    - \frac{1}{r}  + \frac{\ln(\rho)}{\partial r} 
  \right)
  +\frac{r}{u_{\rm kep}^2}F_{\rm lorentz} .
\end{equation}
It follows from the last equations, the gas may reach purely Keplerian
rotation when the density profile is locally changed to $\rho \propto
r$ {\it and} the Lorentz forces are negligibly weak.

The vertical profiles of turbulent velocity dispersion and the mean
radial drift velocity $\Delta U=u_{\phi}-u_{\rm kepl}$ are plotted in
Fig.~\ref{rot1} ($\rm M_{IR}$) and Fig.~\ref{rot2} ($\rm M_{IR6}$).
The root-mean-square turbulent velocities can be described with the
same shape of the vertical profile both in the active and in the dead
zone.  The turbulent velocity dispersion reaches 0.4 in the
corona. This agrees with the result of Fromang \& Nelson (2006).  The
top left panels in Figs.~\ref{rot1} and Fig.~\ref{rot2} show the
root-mean-square turbulent velocities averaged separately for active
(A) and for dead (D) zones. It is interesting to note, that the levels
of $\sqrt{<u_{\Theta}^2/c_{s}^2>}$, $\sqrt{<u_{r}^2/c_{s}^2>}$ at the
midplane are not very different for the two zones.  Midplane values
vary from 0.03 for $\sqrt{<u_{\Theta}^2/c_{s}^2>}$ to 0.16 for
$\sqrt{<u_{\phi}^2/c_{s}^2>}$.  As expected, the perturbations of
rotational profile are higher in the active zone.  The radial
dependence of $\sqrt{<u_{\Theta}^2/c_{s}^2>}$,
$\sqrt{<u_{r}^2/c_{s}^2>}$ shows that the turbulent dispersion is
significantly reduced within the pressure bump at the inner edge of
the dead zone (Fig.~\ref{rot1} and Fig.~\ref{rot2}, bottom right).

The time-average of $\Delta U$ shows that the super-Keplerian rotation
at the density rings is a long-lasting effect.  Relatively large dust
particles (i.e. for Stokes number $\rm St=1$) will migrate outwards
with the velocity $v_{r}=\Delta U$ \citep{kla01}.  The areas of
outward migration are up to $0.5AU$ broad and outward velocities reach
0.15 km/s (Fig.~\ref{rot1} and Fig.~\ref{rot2}, top right).

There are certain similarities between the density bump and rings we
found and the zonal flows described in \citet{joh09}.  The enhancement
in density has been reaching 10 pressure scale heights both in large
local-box simulations and in the global model of \citet{lyr08}.  We
estimate the rings to be roughly 1.5 AU broad, what gives maximum 6
pressure scale heights.  We show the radial correlations of the
following parameters: $\Delta U=u_{\phi}- U_{\rm Kepl}$ for various
heights in the disk, $\delta \rho/\rho_{0}=(\rho-\rho_{0})/\rho_{0}$,
magnetic pressure $B^2/B_{0}^2$, and Maxwell stress $B_{r}B_{\phi}/
P_{\rm init}$ (Fig.~\ref{rot1}, Fig.~\ref{rot2}, bottom left).  The
gas around the rings of enhanced density within the active zone in
$\rm M_{IR6}$ and $\rm M_{IR}$ models shows similar turbulent
properties to the zonal flows and corresponding density maximum as in
\citet{joh09}: The magnetic pressure and Maxwell stress are strongest
where the density minima are excavated, the maximal $\Delta U$ is
half-phase shifted.  The maximal deviations from Kepler velocity have
same amplitude as in zonal flows in \citet{joh09}, but the relative
amplitude of density $\delta \rho/\rho_{0}=(\rho-\rho_{0})/\rho_{0}$
is ten times larger in our models.  The maxima of magnetic pressure
$B^2/B_{0}^2$ and Maxwell stress $B_{r}B_{\phi}/ P_{\rm init}$ are
located between the rings and are about ten times stronger compared to
the values in \citet{joh09}.

In addition, the pressure bump at the ionization thresholds has
significantly less turbulence; the $u_{\Theta}^2/c_{\rm s}^2$ has a
value only about 0.05 and not 0.5 as outside of the density bump. In
the case of $\rm M_{IR}$, the density ring in the active zone is a
'calm place' as well.  Model $\rm M_{IR6}$ keeps the butterfly pattern
of the azimuthal fields much longer.  This is the reason why the
turbulent velocities decrease in the density rings weaker in the case
of $\rm M_{IR6}$, compared to $\rm M_{IR}$.

%%%%%%%%%%%%%%%%%%%%%%%%%%%%%%%%%%%%%%%%%%%%%%%%%%%%%%%%%%%%%%%%%%%%%%%%%%%%%%%%

\section{Synthesis: Connection Between Pressure Maxima and
  `Butterfly' Structures }

\subsection{Density Rings in the MRI-Active Region}

In order to provide the comparison with previous global simulations
\citep{fro06}, we have done the fully MRI-active disk model.  The
turbulent and magnetic properties of the fully MRI-active disk model
$\rm M_{IDEAL}$ are very close to those presented in \citet{fro06} for
run S4.  The toroidal magnetic field is expelled from the midplane
into the upper disk layers within the linear stage of MRI turbulence.
The peak value of the volume-integrated turbulent $\alpha$ stress is
reaching 0.019 in our model $\rm M_{IDEAL}$ (Fig.~\ref{energ}) and
0.013 in the S4 model \citep{fro06}.  At the end of the simulation,
the turbulent $\alpha$ stress is $0.005$ in gas pressure units.

In our simulations, the MRI turbulence evolves through three stages:
(a) linear growth, (b) oscillatory saturation regime and (c)
non-oscillating steady state.  The stage (b) is best to observe if the
dead zone is included. The oscillations are regular and can be
registered in total magnetic energy and in turbulent $\alpha$ stress
time evolution.  The dominating azimuthal magnetic field component in
the MRI-active zone switches its sign with a period of about 150 years
or 30 local orbits, and within the radial extent of 1.5 AU.  The sign
reversals of the azimuthal magnetic field with respect to the
midplane, known as butterfly diagram, have also been observed in local
shearing-box simulations of the stratified disk \citep{tur07,joh09}.
In case of the global simulation (model $\rm M_{IDEAL}$), averaging
over the whole eight AU of the MRI-active domain brings a very
irregular butterfly picture.  Numerous reversals of the azimuthal
magnetic field along the radial extent lead to seemingly irregular
peaks in $\alpha(r)$ and in the magnetic energy.

We observe in our models, that the reversals in $B_{\phi}$ come along
with the density rings.  Thus, it is important to understand what
causes radial and temporal reversals in the magnetic field, and what
determines a period.  We observe that the amplitude of the
oscillations in magnetic energy is higher, if we increase the radial
extent of the active zone (models $\rm M_{IR}$ and $\rm M_{IR6}$).  In
addition, it is important to know how long the 'butterfly' structure
can survive.  The magnetic field is oscillating over the whole
duration of the local-box simulations.  In our global runs, the
life-time of the oscillating stage (b) seems to depend on the extent
of the MRI-active zone (Fig.~\ref{energ}).  The thickness of the dead
zone also influences the life-time of the oscillatory regime (b), as
we found from a comparison of magnetic energy curves for models $\rm
M_{IR}$ and $\rm M_{ID}$.  The MRI waves in active layers above and
below the dead zone are interacting with the turbulent magnetic field
inside the threshold radius.  If the layers above and below the dead
zone are as thin as in our model $\rm M_{ID}$, the turbulent $\alpha$
stress in the active zone appears to be lower, the same applies to the
total magnetic energy.  The oscillations of the azimuthal magnetic
field have ceased after 400 years, and after 600 years in the case of
thinner dead zone (model $\rm M_{IR}$, Fig.~\ref{energ}).  The
turbulent $\alpha$ stress is roughly constant with radius during the
non-oscillating steady stage (c).

The inner radial boundary with its resistive buffer (${2\rm AU}<r<2.5
\rm AU$) and the inner edge of the dead zone $r_{\rm th}$ enclose the
MRI-active zone in our models $\rm M_{IR}$, $\rm M_{ID}$ and $\rm
M_{IR6}$.  We observe a butterfly diagram in the MRI-active zone of
our models, when plotting $\langle{B_{\phi}(z)\rangle} \propto t$.
There are no oscillations of $\langle{B_{\phi}(z)\rangle}$ in or
around of the dead zone.  We obtain the clearest oscillations of
$B_{\phi}$ at the midplane of active zone, $|\Theta-\pi/2|\leq{} c_{0}
$. The $B_{\phi}$-sign reversal happens every 150 years and within
$2.5\rm AU<r<4 \rm AU$ (models $\rm M_{IR}$, $\rm M_{IR6}$) and $4\rm
AU<r<5.5 \rm AU$ ($\rm M_{IR6}$), i.e. within every $1.5$ AU.  Those
reversals are stretched along the line of the local orbit in $(r,t)$
space.

The rings of enhanced density appear already at $t=150$ years, roughly
at the time when the linear AMRI breaks into a nonlinear regime and
the 'butterfly' is initiated.  During the oscillatory stage (b) of the
MRI evolution, there is a radial dependence in turbulent $\alpha$
stresses according to $\alpha(r)\propto r^{-2}$.  At the inner radii,
the $\alpha$ stress is high and it leads to the effect of fast local
excavation of the density and accumulation of it at some outer radius.
The radial reversals of the azimuthal magnetic field are aligned with
the rings of enhanced density.  The radial location of magnetic field
reversals remains constant over hundreds of years.  At the same
location, we find the stripes of weakest Maxwell stress $\sim 10^{-7}$
(Fig. 11).  As soon as the ring of density is created at a certain
location, there is a corresponding change in the rotation and in the
shear.  On the one hand, over-density leads to the local $P_{\rm
  mag}/P_{\rm gas}$ relation which can be too low to excite AMRI.  On
the other hand, the change in the rotation reduces the shear and it
leads to local stabilization of MRI within the density ring.  The
consequence is that the rings of enhanced density are less turbulent
compared to the density minima between the rings.  In contrast to the
local-box studies \citep{joh09,gua09}, there is more then one density
ring forming in the case of a longish MRI-active zone.  The merging
between rings is possible.  In the $\rm M_{IDEAL}$ model, we observe
the appearance of three weak density rings.

\subsection{Pressure Maximum at the Inner Edge of the Dead Zone}

The formation of the pressure bump at the dead zone edge is not
directly correlated with the 'butterfly' magnetic structures within
the active zone.  The strength of the turbulent viscosity in the
MRI-active zone determines the time scale to form such a pressure
bump.  The extent of the MRI-active zone has an effect on the value of
the total turbulent $\alpha$ stress in the active zone.  As it follows
from Table 1, the turbulence in models $\rm M_{IDEAL}$ and $\rm
M_{IR6}$ provide the largest stresses of $5.3\cdot{}10^{-3}$ and
$2.9\cdot{}10^{-3}$, followed by $\rm M_{IR}$ with $1.9\cdot{}10^{-3}$
and $\rm M_{ID}$ with $1.6\cdot{}10^{-3}$. The radial extent of the
active zone was reduced from 8AU to 4 AU and to 2AU.  This sequence is
consistent with the results of \citet{gua09,bod08}, where a similar
decrease of total alpha stress is demonstrated when the size of the
local box is reduced.

The jump in the turbulent viscosity is responsible for the formation
of a pressure trap at the dead zone edge. This jump is strongest at
the midplane.  The intensity of MRI-driven turbulence grows strongly
with the disk height.  The locally calculated stresses are different
for each pressure scale height layer: there is up to 1 order of
magnitude difference when comparing the values for midplane and top
$(|\Theta-\pi/2|>3c_{0})$ layers of the active zone.  As expected, the
midplane value of Maxwell stress decreases drastically in the dead
zone, but the Reynolds stress does not 'feel' the presence of the dead
zone. We observe only a slight bump in the Reynolds stress at the
location of the density maximum.  Within the dead zone, there are
spiral density waves propagating from the inner edge outwards.  The
vertical velocity dispersion is non-zero as well. We conclude that the
resulting $\alpha$ of about $10^{-3}$ is due to the waves pumped
vertically from the active layers and radially from the active zone
through the threshold radius.  For the disk evolution, it is important
to have a significant residual $\alpha$ in the dead zone in order to
reach the quasi-steady state without getting unstable (i.e., a
gravitational instability of the density ring).

%%%%%%%%%%%%%%%%%%%%%%%%%%%%%%%%%%%%%%%%%%%%%%%%%%%%%%%

\section{Conclusions}

We have presented the results of the first global non-ideal 3D MHD
simulations of stratified protoplanetary disks.  The domain spans the
transition from the MRI-active region near the star to the dead zone
at greater distances.  The main results are as follows.
\begin{itemize}
\item[-] As suggested in \citet{kre08}, the height-averaged accretion
  stress shows a smooth radial transition across the dead zone edge.
  The stress peaks well off the midplane at
  $2c_{0}<|\Theta-\pi/2|<3c_{0}$.  Consequently, averaging over the
  full disk thickness yields only a mild jump, despite the steepness
  of the midplane radial stress gradient.
\item[-] Weak accretion flows within the dead zone are driven mainly
  by Reynolds stresses.  Spiral density waves propagating horizontally
  produce non-zero velocities and angular momentum transport near the
  midplane, apparently with little associated mixing.  The dead zone
  midplane vertical velocity dispersion $|u_{\Theta}^{'}|$ is $0.03$
  times the sound speed, two to three times less than the radial and
  azimuthal components (Fig.~\ref{rot2}).  The waves are pumped both
  by the active region near the star and by the active layers above
  and below the dead zone.  In model $\rm M_{ID}$ where the dead zone
  is very thick, the pumping from the surface layers is weak and the
  stress falls steeply with radius within the dead zone
  (Fig.~\ref{alp1}).
\item[-] The excavation of gas from the active region during the
  linear growth and after the saturation of the MRI leads to the
  creation of a steady local radial gas pressure maximum near the dead
  zone edge, and to the formation of dense rings within the active
  region, resembling the zonal flows described in \citet{joh09}
  (Fig.~\ref{surd1}, Fig.~\ref{dens}).  Super-Keplerian rotation is
  observed where the radial pressure gradient is positive.  The
  corresponding outward radial drift speeds for bodies of unit Stokes
  number can exceed 10\% of the sound speed.  The pressure maxima are
  thus likely locations for the concentration of solid particles.
\item[-] The turbulent velocity dispersion, magnetic pressure and
  Maxwell stresses all are greatest in the density minima between the
  rings.  The dense rings are `quiet' locations where the turbulence
  is substantially weaker.  Such an environment may be helpful for the
  growth of larger bodies.
\item[-] The rings within the active zone sometimes move about,
  leading to mergers.  In contrast, the bump at the dead zone inner
  edge is stationary.  Planetary embryos with masses in the type~I
  migration regime can be retained at the dead zone inner edge as
  proposed by \citet{ida08a,sch09}.
\end{itemize}

\section{Outlook}

The causes of the magnetic field oscillations appearing in the
butterfly diagram remain to be clarified.  In particular, it is not
known what processes drive the quasi-periodic reversals in the
azimuthal magnetic fields shown in figure~\ref{fluxb3}.

While we have used a fixed magnetic diffusivity distribution, the
degree of ionization will in fact change as the disk evolves.  Annuli
of increasing surface density will absorb the ionizing stellar X-rays
and interstellar cosmic rays further from the midplane, and will have
higher recombination rates due to the greater densities.  These
effects will likely mean contrasts in the strength of the turbulence
between the rings and inter-ring regions, even greater than those
observed in our calculations.  Stronger magnetic fields and higher
mass flow rates in the inter-ring regions could lead in turn to growth
in the surface density contrast over time, analogous to the viscous
instability of the $\alpha$-model in the radiation-pressure dominated
regime \citep{lig74,pir78}. 
%(Lightman \& Eardley 1974; Piran 1978).

The resistivity can show another kind of time variation near the
boundary of the thermally-ionized region.  Changes in the temperature
can alter the strength of the turbulence and thus the heating rate.
In this way, radial heat transport can activate previously dead gas
\citep{wue05,wue06}.  Radial oscillations of the ionization front may
be the consequence.  Overall, due to the fixed magnetic diffusivity,
it is probable that our models underestimate the evolution resulting
from the ionization thresholds.

Owing to the effects of the radial boundary conditions, our global
simulations have limitations for measuring quantities such as the
accretion rate and the mean radial velocity.  Fixing the rate of flow
across the outer radial boundary is a promising avenue for further
exploration in this direction.

%%%%%%%%%%%%%%%%%%%%%%%%%%%%%%%%%%%%%%%%%%%%%%%%%%%%%%%%%%%%%%%%%%%%%

\paragraph{Acknowledgments}
N. Dzyurkevich acknowledges the support of the Deutsches Zentrum f\"ur
Luft- und Raumfahrt (DLR).  N. Dzyurkevich, M. Flock and H. Klahr were
supported in part by the Deutsche Forschungsgemeinschaft (DFG) through
Forschergruppe 759, ``The Formation of Planets: The Critical First
Growth Phase''.  The participation of N. J. Turner was made possible
by the NASA Solar Systems Origins program under grant 07-SSO07-0044,
and by the Alexander von Humboldt Foundation through a Fellowship for
Experienced Researchers.  The parallel computations were performed on
the PIA cluster of the Max Planck Institute for Astronomy Heidelberg,
located at the computing center of the Max Planck Society in Garching.

\bibliographystyle{aa}

\bibliography{12834nsd}

\end{document}